\documentclass[fleqn,10pt]{wlscirep}
\usepackage[utf8]{inputenc}
\usepackage[T1]{fontenc}
\usepackage{amsfonts, amsmath, amssymb, amsthm, color, bm}
\usepackage{bbold}
\usepackage{braket}
\usepackage{multicol, multirow}
\usepackage{algorithmic}

\usepackage{subfig}
\DeclareRobustCommand{\times}{\fontfamily{artimes}\selectfont}
\usepackage{braket}
\usepackage[ruled,vlined]{algorithm2e}
\newcommand{\etal}{\textit{et al.}}
\usepackage{nicefrac}

\title{Solving Linear Systems on Quantum Hardware with Hybrid HHL$^{++}$}

\author[1,*]{Romina Yalovetzky}
\author[1]{Pierre Minssen}
\author[1]{Dylan Herman}
\author[1]{Marco Pistoia}
\affil[1]{Global Technology Applied Research, JPMorganChase, New York, NY 10017, USA}

\affil[*]{romina.yalovetzky@jpmorgan.com}

\begin{abstract}

The limited capabilities of current quantum hardware significantly constrain the scale of experimental demonstrations of most quantum algorithmic primitives. This makes it challenging to perform benchmarking of the current hardware using useful quantum algorithms, i.e., application-oriented benchmarking. In particular, the Harrow-Hassidim-Lloyd (HHL) algorithm is a critical quantum linear algebra primitive, but the majority of the components of HHL are far out of the reach of noisy intermediate-scale quantum devices, which has led to the proposal of hybrid classical-quantum variants. The goal of this work is to further bridge the gap between proposed near-term friendly implementations of HHL and the kinds of quantum circuits that can be executed on noisy hardware. Our proposal adds to the existing literature of hybrid quantum algorithms for linear algebra that are more compatible with the current scale of quantum devices. Specifically, we propose two modifications to the Hybrid HHL algorithm proposed by Lee \etal~\cite{lee2019hybrid} leading to our algorithm Hybrid HHL$^{++}$: (1) propose a novel algorithm for determining a scaling factor for the linear system matrix that maximizes the utility of the amount of ancillary qubits allocated to the phase estimation component of HHL, and (2) introduce a heuristic for compressing the HHL circuit. We demonstrate the efficacy of our work by running our modified Hybrid HHL on Quantinuum System Model H-series trapped-ion quantum computers to solve different problem instances of small-scale portfolio optimization problems, leading to the largest experimental demonstrations of HHL for an application to date. 

\end{abstract}
\begin{document}

\flushbottom
\maketitle

\section{Introduction}

\label{section_introduction}

Recent advances in quantum hardware~\cite{Arute2019,Wu2021,Madsen2022} open the path for experimentation with quantum algorithms. This has led to a rise in \emph{application-oriented} benchmarking of various commercial quantum devices \cite{lubinski2023applicationoriented, lubinski2024quantum}. Due to the amount of noise present in the Noisy Intermediate-Scale Quantum (NISQ) hardware~\cite{preskill2018quantum}, it might be necessary to introduce modifications to the quantum algorithms being tested to make them amenable to such devices. We are not currently at the scale where asymptotics play a role, and thus we can swap out some of the components of these algorithms that provide the promised asymptotic speedup for ones that are easier to use for testing the current hardware. These optimizations can be removed once the hardware becomes fault-tolerant. The focus of this work will be on introducing such optimizations to the quantum algorithm for linear systems, HHL, and executing the modified algorithm on a trapped-ion quantum device.

The HHL algorithm was introduced by Harrow, Hassidim and Lloyd~\cite{harrow2009} for solving a quantum version of the linear systems problem and is a key component in various algorithms and applications. 
% Quantum Linear Systems Problem (QLSP). A linear system is of the form $A \vec{x} = \vec{b}$, where $A \in \mathbb{C}^{N \times N}$ and $\vec{x}, \vec{b} \in \mathbb{C}^N$, thereby returning the quantum state $\ket{x}$ corresponding, up to a normalization factor, to the solution of the linear system. This quantum algorithm has many applications \cite{}. 
In particular, HHL has been proposed as a possible solver for a specific portfolio-management problem~\cite{rebentrost2018quantum}, known as mean-variance portfolio optimization~\cite{markowitzmpt1952}. However, HHL is known to be cumbersome to deploy~\cite{aaronson2015read}, especially on NISQ hardware. 
Recently, hybrid classical-quantum versions of quantum linear systems solvers, such as variational approaches \cite{bravoprieto2020variational, huang2019near} and a hybrid version of HHL \cite{lee2019hybrid, zhang2022improved} have been developed to enable near-term experimentation with quantum linear algebra techniques. 

The Hybrid HHL, originally developed by Lee \etal~\cite{lee2019hybrid}, was focused solely on reducing the complexity of the eigenvalue inversion step, at the cost of producing an algorithm that does not have an asymptotic speedup over classical approaches. This is an example of the types of modifications mentioned earlier. This approach avoids, currently, infeasible quantum arithmetic and exponentially deep uniformly controlled rotation gates \cite{mottonen2004transformation}. However, we believe that further optimizations of the Hybrid HHL are still possible, especially given newly available quantum hardware features. The culmination of the techniques we introduce will be the Hybrid HHL$^{++}$, which we believe provides a viable near-term application-oriented benchmark, which we show by applying it to small-scale portfolio optimization problems.

To highlight, our work makes the following novel contributions:

\begin{enumerate}

    %\item The integration of semiclassical QPE into the Hybrid HHL framework introduced by Lee \etal~This procedure only requires one ancillary qubit to estimate eigenvalues to arbitrary precision. %We benchmark the performance of this  component on the trapped-ion Quantinuum System Model H$1$. and we do a comparative analysis of the required two-qubit gate and ancillary qubit count. %The benchmark of the performance of a version of QPE that makes use of the DQCs, the semiclassical QPE, on real quantum hardware---the trapped-ion Quantinuum System Model H$1$.

    \item Introduction of the Hybrid HHL$^{++}$ algorithm, which incorporates two new features into the exisiting Hybrid HHL pipeline:
    \begin{enumerate}
        \item an efficient and verifiable algorithmic procedure for determining a factor to scale the system matrix by, which allows for resolving the eigenvalues with significantly higher accuracy and making better use of the allocated number of qubits,%, in the output distribution. This allows the Hybrid HHL algorithm to output a quantum state with a high fidelity. 
    
        \item and a generic procedure for reducing the complexity of Hybrid HHL circuit, in terms of both the number of qubits and the number of rotations used by the eigenvalue inversion circuit. This protocol makes use of the semiclassical quantum phase estimation \cite{semiclassicalqft} and leverages newly available hardware features, such as conditional logic and mid-circuit measurements.

    \end{enumerate}

   \item To the best of our knowledge, we demonstrate the largest-to-date execution \cite{lee2019hybrid, zhang2019simplified, zhang2022improved, lubinski2024quantum} of HHL with a two-qubit gate depth up to 291 on real quantum hardware---the Quantinuum H-series trapped-ion devices---to solve portfolio-optimization problems with \mbox{S\&P 500} assets.

\end{enumerate}

\noindent \textbf{Paper Organization}
The remainder of this paper is organized as follows. Section~\ref{sec:rev_of_hybrid} reviews both HHL and Hybrid HHL, Section~\ref{sec:hybrid_hhl_plusplus} introduces the novel techniques for enhancing Hybrid HHL culminating in the Hybrid HHL$^{++}$ algorithm. Section \ref{sec:results} presents the results of our experiments from simulation and the Quantinuum H-series devices and noisy emulators. Section \ref{section:scale} presents an analysis of future hardware demonstrations by simulating different levels of noise on the quantum hardware and increasing the problem size. Finally, Section~\ref{sec:Conclusions} summarizes the results of this work, and concludes the article.

% The remainder of this paper is organized as follows. Section~\ref{section:eigenvalue_inversion} describes the eigenvalue inversion component of HHL and the different approaches that have been proposed in the literature. Section~\ref{section:Techniques for Enhancing Hybrid HHL} introduces the novel techniques for enhancing Hybrid HHL. In particular, Section \ref{section:semi_classical_qpe} introduces semiclassical QPE, and shows benchmarks on quantum hardware, the trapped-ion Quantinuum System Model H$1$. This section also explains how the semiclassical QPE is integrated to the Hybrid HHL to estimate the relevant eigenvalues. Section~\ref{sec:evol_time_optimization} introduces the algorithmic contributions, focusing on its novelties. It also contains results obtained from experiments on hardware showing how these algorithms allow for estimating the eigenvalues accurately. Section~\ref{section:Classical-Quantum Hybrid HHL with Quantum Conditional Logic} describes a known formulation of portfolio optimization as a QLSP, which makes it compatible with HHL and it presents a detailed comparative analysis of the hardware demonstrations of the end-to-end Hybrid HHL with dynamic quantum circuits to solve this particular problem. Section~\ref{section:approaches} discusses different approaches for the eigenvalue inversion in terms of hardware features. Finally, Section~\ref{sec:Conclusions} summarizes the results of this work, and concludes the article.

\section{Review of HHL and Hybrid HHL}

\label{sec:rev_of_hybrid}
The original HHL algorithm was proposed to solve the following quantum analog of a linear system. Given oracle access to the entries of an $2^n \times 2^n$ Hermitian matrix $\mathbf{A}$ and an $n$-qubit quantum state $\ket{\mathbf{b}}$ prepare the state, \begin{align}
    \ket{\mathbf{x}} = \frac{\mathbf{A}^{+}\ket{\mathbf{b}}}{\lVert \mathbf{A}^{+}\ket{\mathbf{b}}\Vert_{2}},
\end{align}
where $\mathbf{A}^{+}$ is the Moore-Penrose pseudoinverse of $\mathbf{A}$. The restrictions on $\mathbf{A}$ being square and Hermitian can be relaxed by dilation. The HHL procedure to prepare $\ket{\mathbf{x}}$ is roughly as follows:
\begin{itemize}
    \item \textbf{Step 1:} Perform $m$-qubit quantum phase estimation (QPE) to the unitary $e^{2\pi i\gamma\mathbf{A}}$ (obtained via a Hamiltonian simulation algorithm) and apply it to the state $\ket{\mathbf{b}}$ to obtain:
    \begin{align}
\sum_{\lambda}c_{\lambda}\ket{\psi_{\lambda}}\ket{2^m\gamma\lambda},
    \end{align}
    where $\ket{\psi_{\lambda}}$ is an eigenstate of $\mathbf{A}$ with eigenvalue $\lambda$, $c_{\lambda} = \langle \mathbf{b}| \psi_{\lambda}\rangle$. For simplicity, we  assume that there is a known $\gamma$ s.t. $\forall \lambda$ s.t. $\lvert c_{\lambda}\rvert > 0$:
    \begin{align}   
    2^m\gamma\lambda \in \{-2^{m-1}, \dots, 0, \dots, 2^{m-1}-1\} =: \mathcal{B}_{m}, 
    \end{align} represented using two's complement (i.e. that phase estimation can be performed exactly).
    \item \textbf{Step 2:} Perform the following unitary (called \emph{eigenvalue inversion}) for $\lambda \neq 0$:
    \begin{align}
        |0\rangle|2^m\gamma\lambda\rangle \mapsto \frac{C}{\lambda}|0\rangle|2^m\gamma\lambda\rangle + \sqrt{1 - \frac{C^2}{\lambda^2}}|1\rangle|2^m\gamma\lambda\rangle,
    \end{align}
    where $C$ is chosen so that $\lvert C/\lambda\rvert \leq 1$ for all $\lambda$.
    \item \textbf{Step 3:} Uncomputing and post-selecting on $\ket{0}$ gives
    \begin{align}
        \frac{1}{\mathcal{N}}\sum_{\lambda}\frac{c_{\lambda}}{\lambda}\ket{\psi_{\lambda}} = \ket{\mathbf{x}}.
    \end{align}
\end{itemize}
There can be an exponential quantum speedup in terms of the number of  queries to the oracles for $\mathbf{A}$ and $\ket{\mathbf{b}}$ when $\mathbf{A}$ is sufficiently sparse and well-conditioned. However, this algorithm is still not feasible on near-term devices. Our work is focused on finding more near-term friendly versions of the original HHL that, at the expense of losing the HHL speedup, can be implemented on current hardware and used as an application-oriented benchmark. The algorithm we will be focused on improving, the Hybrid HHL, is one such approach to making HHL compatible with small quantum devices. In addition, due to Hamiltonian simulation being generally infeasible on current devices for an arbitrary Hermitian $\mathbf{A}$ and $2^n$ being small enough that a classical computer can solve the problem, we simply fallback on the following hybrid simulation procedure: compute the unitary matrix $\mathbf{U}=e^{2\pi i\gamma\mathbf{A}}$ on a classical device and use standard quantum circuit packages, e.g., Qiskit \cite{aleksandrowicz2019qiskit}, to decompose $\mathbf{U}$ into basis gates. Specifically, Qiskit makes use of product formulas to approximate the simulation.

The key step that distinguishes Hybrid HHL, and removes any potential speedup (besides the hybrid Hamiltonian simulation procedure), is that we obtain classical  access to the eigenvalues of $\mathbf{A}$ (via either sampling from the QPE output distribution or a classical eigensolver). This allows us to implement the eigenvalue inversion (step 2 in the procedure above) by applying a single multi-controlled rotation for each eigenvalue $\lambda$. While utilizing QPE instead of a classical eigensolver to obtain the eigenvalues provides no computational benefit in the current setting, Hybrid HHL and our approach opt to utilize QPE instead so as to simulate the HHL algorithm more closely. The benefits of estimating the eigenvalues a priori in HHL for demonstrations on small hardware are two-fold: (1) it removes the need to use resource-heavy quantum arithmetic and (2) the number of controlled rotations in the eigenvalue inversion circuit only scales with the number of distinct eigenvalues estimated. One alternative to the second point is known as the uniformly controlled rotation gate, which performs a multi-controlled rotation for every basis state that $\ket{2^m\gamma\lambda}$ could be, exponentially scaling with the number of ancillary qubits.

The following is a summary of the two main components of Hybrid HHL:
\begin{enumerate}
    \item Eigenvalue Estimation
    \begin{enumerate}
    \item Perform step 1 of the standard HHL algorithm using the hybrid Hamiltonian simulation procedure described earlier.
    \item Measure the register containing $|2^m\gamma \lambda\rangle$ to obtain a probability distribution of the eigenvalue estimates such that $\lvert c_{\lambda}\rvert > 0$ or is above some threshold value.
    \end{enumerate}
    \item Compute $|\mathbf{x} \rangle$
    \begin{enumerate}
    \item  Perform step 1 of the standard HHL again on a fresh copy of $|\mathbf{b}\rangle$.
    \item For each observed $\lambda$ in step 1, apply the following  multi-controlled  rotation $\mathsf{R}_{\mathsf{Y}}$:
    \begin{align}
|2^m\gamma\lambda\rangle\langle2^m\gamma\lambda| \otimes \mathsf{R}_{\mathsf{Y}}(2\sin^{-1}(C/\lambda)) + [\mathbb{I} - |2^m\gamma\lambda\rangle\langle2^m\gamma\lambda|]  \otimes \mathbb{I}.
    \end{align}
    \item Uncompute and post select on $|0\rangle$ as in standard HHL.
    \end{enumerate}
\end{enumerate}

Lee \etal~ also proposed an additional optimization that allows the number of controlled rotations in the eigenvalue inversion to be reduced further under additional assumptions. We will address this in a later section and present a potential improvement. While the hybrid version of HHL is significantly more near-term compatible than the original version, there are still additional optimizations that can be made. As mentioned in the introduction, our work presents two additional modifications, which we consolidate into the Hybrid HHL$^{++}$ algorithm.

\section{Hybrid HHL$^{++}$}
\label{sec:hybrid_hhl_plusplus}

\begin{figure*}[!h]
    \centering
    \includegraphics[width=0.8\linewidth]{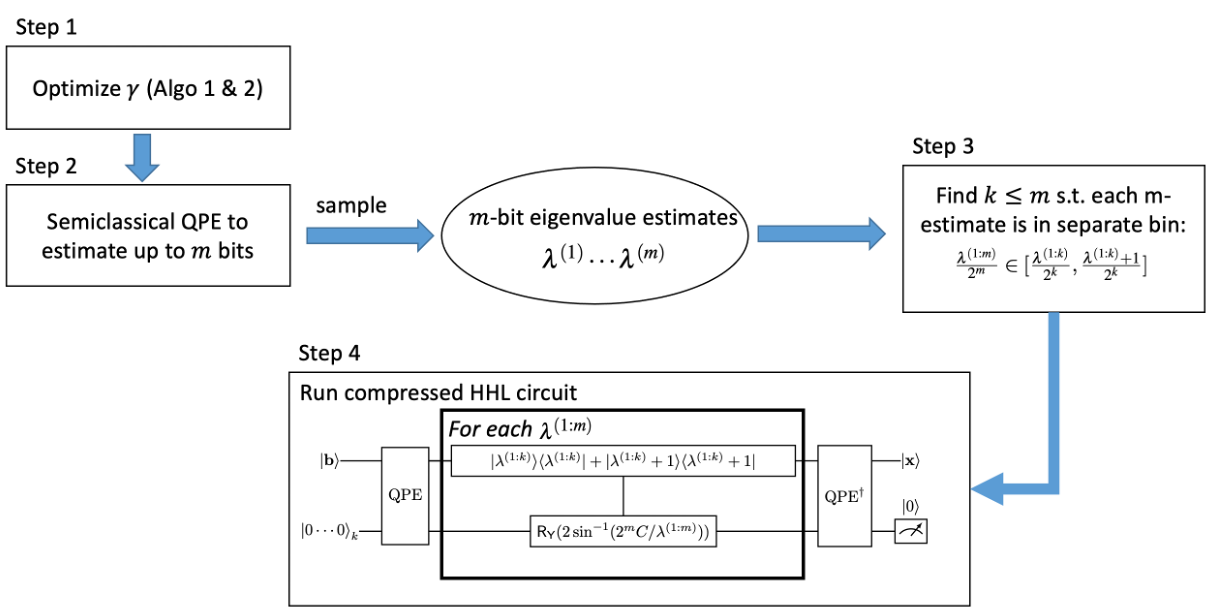}
\caption{End-to-End Flow of Hybrid HHL$^{++}$: highlights the integration of the two proposed modifications, the spectral scaling algorithm (Section \ref{sec:spectral_scaling}) in Step 1 and the HHL circuit compression (Section \ref{section:eigenvalue_inversion}) techniques in Step 3 and Step 4, into the original Hybrid HHL flow.}
% (a) Semiclassical QPE is used to construct a distribution over the estimates of the relevant eigenvalues with $m$-bit precision. The optimal scaling parameter, $\gamma$, for QPE is determined by  running algorithms \ref{alg:time_evolution_optimization} and \ref{alg:inital_guess_checker}.
% (b) Classical post-processing is performed on the resulting histogram to obtain the estimates of the $d$ relevant eigenvalues.
% (c) The $m$-bit estimates, $\{\tilde{\lambda}_i\}_{i=0}^{d-1}$, obtained in (b), are used to determine rotation angles $\{2\arcsin(\nicefrac{C}{\tilde{\lambda}_i})\}_{i=0}^{d-1}$ for the eigenvalue inversion circuit. %The estimates are also mapped to a smaller number of bits, $r$. Each rotation is conditioned on its corresponding $r$-bit estimate.
% (d) The standard HHL procedure is executed, but it uses the circuit constructed in (c) for the eigenvalue inversion step. \textcolor{red}{TODO: update to reflect changes}}
\label{diagram:algorithm}
\end{figure*}

In this section we present our improved Hybrid HHL procedure, which enables larger-scale experimental demonstrations of quantum linear algebra than prior work. As mentioned earlier, Hybrid HHL$^{++}$ differs from the proposal of Lee \etal~ in two ways: the spectral scaling algorithm (Section \ref{sec:spectral_scaling}) and the heuristic for compressing the HHL circuit (Section \ref{section:eigenvalue_inversion}). The following two subsections will describe these modifications in detail. Figure \ref{diagram:algorithm} shows the end-to-end flow combining all of the techniques introduced.

%First, in Step 1 we optimize the scaling parameter $\gamma$ with the Algorithms 1 and 2 that will be introduced in Section \ref{sec:spectral_scaling}. Then, in Step 2, with this optimal $\gamma_{op}$ we execute the semiclassical QPE to estimate the eigenvalues up to $m$-bits. We then sample from this distribution and we create the set of the $m$-bit eigenvalue estimates. In Step 3, we perform the heuristic method we will introduce in Section \ref{section:eigenvalue_inversion} to find $k\leq m$ such that each $m$-bit estimate is in a separate bin. Finally, in Step 4, we construct the compressed eigenvalue inversion circuit. We map the $m$-bit bitstrings to $k$-bit bitstrings to control the rotations on and we calculate the rotation angles with the $m$-bit eigenvalue estimates. Then, we insert this component to the HHL circuit, resulting in the compressed HHL circuit ready to be run.}

\subsection{Spectral Scaling Algorithm for Optimizing $\gamma$}
\label{sec:spectral_scaling}
As hinted at in the presentation of the HHL and Hybrid HHL algorithms in the previous section, the choice of $\gamma$, the parameter to scale $\mathbf{A}$ by, is critical to the performance of phase estimation. Given that the eigenvalues of $\mathbf{A}$ could be irrational or rational numbers with incompatible denominators, there may be no such that $\gamma$ that allows for a perfect phase estimation to $m$-bits, i.e., an element of $\mathcal{B}_{m}$. In addition, given that we are working on small quantum devices, it is important to find a $\gamma$ that makes the best use of the allocated $m$ bits of precision. 

The notion of making ``best use'' of the $m$ bits of precision can actually be made somewhat precise. Specifically, the eigenvalues with the largest magnitude should be associated with the elements with largest magnitude of $\mathcal{B}_{m}$.   If we know that the main precondition for QPE to work $\lVert \mathbf{A} \rVert_{2} \leq 1$ is satisfied, and blindly apply phase estimation, then the desired relation between the eigenvalues and elements of $\mathcal{B}_m$ just mentioned may not be satisfied. This can lead to both eigenvalues not being able to be properly resolved and a waste allocated resources. Ideally, we would like to find a $\gamma$ that stretches the  spectrum as to take up as much of  $\mathcal{B}_{m}$ as possible. The goal of this section is to present an efficient and simple procedure for finding such a $\gamma$, which we call the spectral scaling algorithm. Lastly, we would like to emphasize that $\gamma$ only needs to be chosen to ensure high-quality estimation of those $\lambda$ s.t. $\lvert c_{\lambda}\rvert > 0$. This is implicitly accounted for by sampling from the output distribution of QPE with $\mathbf{U}$ applied to the initial state $\ket{\mathbf{b}}$.

% In this section we present the novel algorithmic contributions of this work: the integration of a verifiable algorithm for selecting the optimal scaling parameter to scale the matrix $A$ of the QLSP. This scaling allows for effectively estimating the eigenvalues in the output distribution obtained with the separate QPE component.

% In the foundational HHL article \cite{harrow2009}, $A$ is assumed to have positive eigenvalues in $[\nicefrac{1}{\kappa}, 1]$, where $\kappa$ is the condition number of $A$. The eigenvalues are restricted to this range to account for the periodicity of the imaginary exponential and ensured well-conditioning. In practice, it is necessary to scale $A$ to have a spectrum in this range. However, even under this assumption, we could be wasting qubits to unnecessarily encode values between the largest eigenvalue, $\lambda_\textup{max}$, and $1$. We pursue a more efficient approach, consisting of estimating $\lambda_\textup{max}$ first, and then scaling $A$ by
% $\gamma=\tilde{\lambda}_\textup{max}^{-1}$, so that the maximum eigenvalue of $\gamma A$ is $1$. 

% A benefit of our approach is that it only considers $\lambda_{\textup{max},b}$, the largest eigenvalue in $\Lambda_b$, instead of $\lambda_{\textup{max}}$. For now on, we will only take into account the eigenvalues in $\Lambda_b$ and will not, for example, make a distinction between $\lambda_{\textup{max},b}$ and $\lambda_\textup{max}$. 

\begin{algorithm}[h]
\SetAlgoLined
 Guess an overapproximation $\alpha$ of $\lambda_\textup{max}$\\
 $\gamma := \nicefrac{1}{\alpha}$ \tcp{Initialize scaling parameter}
 $x := 0$\\
 \tcp{At each step, $\gamma*\lambda_\textup{max} \leq 1$}
 
 \While{$x \neq 2^{m} -1$} {
   $p :=$ $n$-bit output distribution of QPE using unitary $e^{ 2\pi i \gamma \mathbf{A}}$ and input state $\ket{b}$\\
   $x :=$ max$\{ j \in \{0, ..., 2^m -1\} | \ p_j > 0, \ p_j \in p\} $ \\
   \tcp{$x$ is an $n$-bit estimation of} \tcp{$2^{m}*\gamma*\lambda_\textup{max}$}
   \eIf{$x=0$}{
   $\gamma := \gamma*2^{m}$\
   }{
   $\gamma := \gamma*\nicefrac{(2^{m}-1)}{x}$ \\
   }
  }
 \KwResult{$\gamma := \tilde{\lambda}_{\textup{max}}^{-1}$, with $\tilde{\lambda}_{\textup{max}}$ $m$-bit estimation of $\lambda_{\textup{max}}$ }
 \caption{Optimize the selection of $\gamma$ using $m$-bit estimations of eigenvalues}
 \label{alg:time_evolution_optimization}
\end{algorithm}

The spectral scaling procedure is highlighted in Algorithm \ref{alg:time_evolution_optimization}, which shows how to optimize the selection of $\gamma$. Given $m$-bits, the optimal value of $\gamma$ returned by the algorithm helps to encode the eigenvalues using all available bits in the output distribution of phase estimation. As a result, this makes it easier to distinguish the eigenvalues from each other and estimate the relevant eigenvalues, %i.e.,, $\Lambda_b$, 
accurately. Without loss of generality, Algorithm \ref{alg:time_evolution_optimization} assumes all the eigenvalues to be positive. We will show shortly how to account for negative eigenvalues.

Algorithm \ref{alg:time_evolution_optimization} starts by guessing an overapproximation of $\lambda_\textup{max}$. We will discuss later how to make this operation rigorous.
The next step consists of iteratively updating the value of $\gamma$ until $\gamma$ converges to the optimal value. During each iteration, Algorithm \ref{alg:time_evolution_optimization} runs QPE to compute the $m$-bit estimates of the eigenvalues of the unitary $\mathbf{U}$ using gamma computed during the previous iteration. Specifically, \mbox{Algorithm \ref{alg:time_evolution_optimization}} post-processes the output distribution to get a new $m$-bit estimation of $2^m \gamma  \lambda_{\textup{max}}$ in order to update $\gamma$ for the next iteration.%\footnote{Any algorithm outputting an $n$-bit estimation of $2^n  \gamma \lambda_{\textup{max}}$ could be used in place of semiclassical QPE. Nevertheless, we chose to use semiclassical QPE, since it is more compatible with NISQ devices, as explained in Section \ref{section:eigenvalue_inversion}.}

The number of iterations to find the optimal $\gamma$ using \mbox{Algorithm \ref{alg:time_evolution_optimization}} is in 
$\Theta\left(\nicefrac{1}{m}\log_2\left(\nicefrac{\alpha}{\lambda_\textup{max}}\right)\right)$. To ensure we do not overestimate $\gamma$ in the process, we could take a conservative approach, which consists of overestimating $x$ while computing it, thus lowering $\nicefrac{(2^{m}-1)}{x}$. In practice, we also want to underestimate $\gamma$ in order to prevent amplitudes of basis states near $2^m$ from dispersing and mixing with basis states near $0$ and causing overflowing. 

Regarding the bit precision, to represent both the largest and smallest eigenvalues, in theory, we need \mbox{$m \geq {\log_2(\nicefrac{\lambda_\textup{max}}{\lambda_{\textup{min}}})} = {\log_2(\kappa)}$}. %However, this requires prior knowledge of $\kappa$. 
In practice, we first run \mbox{Algorithm \ref{alg:time_evolution_optimization}} with an initial precision of $m$ bits. Then, if $\lambda_{\textup{min}}$, estimated via QPE, turns out to be $0$, this indicates that we need higher bit precision to prevent the $0$ state from having significant probability. This is because $\mathbf{A}$ is assumed invertible and consequently, no eigenvalue can be $0$. Thus, we increase $m$ just enough to guarantee that the estimation of $\lambda_{\textup{min}}$ is different from $0$, at which point $m\geq \log_2(\kappa)$.

The correctness of Algorithm \ref{alg:time_evolution_optimization} relies on the fact that $\alpha$ is an overapproximation of $\lambda_\textup{max}$.
One way to achieve this result is to use the Frobenius norm of $A$,  defined as \mbox{$\lVert\mathbf{A}\rVert_\textup{F} := \sqrt{\textup{Tr}(\mathbf{A}^\dagger \mathbf{A})}$}. Indeed, since $\mathbf{A}$ is Hermitian, \mbox{$||\mathbf{A}||_\textup{F} = \sqrt{\sum_{i=0}^{N-1}\lambda_i^2}$}, and so the Frobenius norm of $A$ is a valid overapproximation of $\lambda_\textup{max}$.
However, to make the procedure general, we incorporate a subroutine the finding the overestimate. %the computation of the Frobenius norm has quadratic complexity in $N$. Therefore, it is desirable to find a faster approach for finding a value for $\alpha$ that overapproximates $\lambda_\textup{max}$. 
For this reason, we propose
Algorithm \ref{alg:inital_guess_checker}, which, using one execution of QPE, tests the validity of any initial guess of $\alpha$. %This bypasses the need to compute $||A||_\textup{F}$. 
If the initial guess of $\alpha$ does not satisfy $\alpha \geq \lambda_\textup{max}$, we can retry with a, potentially significantly, larger $\alpha$. In fact, given the logarithmic complexity, in terms of number of iterations, of \mbox{Algorithm \ref{alg:time_evolution_optimization}}, even with low bit precision, such as $n=4$, overestimating $\lambda_\textup{max}$ by a factor of a billion would only require eight iterations before returning the optimal $\gamma$. %In practice, we expect to find $\gamma$ significantly more efficiently by starting with a guess, $\alpha$, than by computing $||A||_\textup{F}$. 

\begin{algorithm}[h!]
\SetAlgoLined
\tcp{Initialize scaling parameter}
 $\Gamma := \nicefrac{1}{2^{m+1} \alpha}$ \\ 
 $p :=$ $m$-bit output distribution of  QPE using unitary $e^{ 2\pi  i\Gamma \mathbf{A}}$ and input state $\ket{b}$\\
 \eIf{$p_0 \neq 1$}{
 Return $\alpha$ is not valid\\
 \tcp{Otherwise all eigenvalues}
 \tcp{estimations would have been $0$}}{
 Return $\alpha$ is valid\\}
 \KwResult{Return if $\alpha$ is an overestimation of $\lambda_{\textup{max}}$}
 \caption{Verify if $\alpha$ is a $n$-bit overestimation of $\lambda_{\textup{max}}$\\
 \textit{\footnotesize{Assumption: at each step, there is at least one eigenvalue of $\gamma \mathbf{A}$ not in $\bigcup\limits_{j\in \mathbb{Z}} \left[j- 2^{-(m+1)},\ j+2^{-(m+1)}\right]$ }}}
 \label{alg:inital_guess_checker}
\end{algorithm}

The idea of Algorithm \ref{alg:inital_guess_checker} is the following. If $\alpha$ is a valid guess, there will be no overflow in the output distribution of QPE applied to $e^{2\pi i\gamma\mathbf{A}}$ with $\gamma = \nicefrac{1}{\alpha}$. Then, performing $m+1$ right bit shifts should reduce all $n$-bit eigenvalue estimates to $0$. To test this, Algorithm \ref{alg:inital_guess_checker} executes QPE using $e^{ 2\pi i\Gamma\mathbf{A}}$ with $\Gamma = \nicefrac{1}{2^{m+1}\alpha}$. On the contrary, if $\alpha$ is not a valid guess, the $m+1$ right bit shifts would not be sufficient to reduce all estimations to $0$. Note, $\gamma$ and $\Gamma$ have the same role in the definition of $\mathbf{U}$ and just help differentiate the two algorithms.

As mentioned before, \mbox{Algorithm \ref{alg:time_evolution_optimization}} assumes positive eigenvalues. To take into account negative eigenvalues, one can
define the maximum eigenvalue using the absolute value, encode negative eigenvalues using two's complement and then replace $2^m$ by $2^{m-1}$ in the update of $\gamma$. Similarly, in \mbox{Algorithm \ref{alg:inital_guess_checker}}, replacing $\nicefrac{1}{2^{m+1}}$ by $\nicefrac{1}{2^{m}}$ in the definition of $\Gamma$ suffices to support negative eigenvalues. Accounting for negative eigenvalues is crucial, since $\mathbf{A}$, in Equation \eqref{linear_system}, may be indefinite.

% We note that Kerenidis and Prakash \cite{Kerenidis2020} have developed an algorithm to $\epsilon$-approximate \mbox{$\eta := \nicefrac{||A||_2}{||A||_\textup{F}}$}, where \mbox{$||A||_2= |\lambda_\textup{max}|$} is the spectral norm of $A$. Thus, their algorithm can also be used to find $ |\lambda_\textup{max}|$ given that $||A||_\textup{F}$ has been previously computed, which is not required for Algorithm \ref{alg:time_evolution_optimization}. As mentioned before, the optimal $\gamma$ is $|\lambda_{\textup{max},b}|^{-1}$ and not $|\lambda_{\textup{max}}|^{-1}$, when given $\ket{b}$ as an initial state. Since Algorithm \ref{alg:time_evolution_optimization} only computes $|\lambda_{\textup{max},b}|^{-1}$, this makes it more suitable for the proposed HHL implementation based on the discussions in Section \ref{section:eigenvalue_inversion}.

% In addition, their algorithm executes standard QPE, which is not exchangeable for semiclassical QPE because of the coherence requirement. In contrast, Algorithm \ref{alg:time_evolution_optimization} runs semiclassical QPE, which for the reasons explained above, is more suitable for NISQ computers.

\subsection{Heuristic for Compressing HHL circuit}
\label{section:eigenvalue_inversion}
In the preliminary eigenvalue estimation step of Hybrid HHL, one could utilize the aforementioned approach of Lee \etal~, where the textbook implementation of phase estimation is used. However, QPE is still difficult to implement on NISQ hardware \cite{mohammadbagherpoor2019experimental} because the number of ancillary qubits grows with the desired bit precision. Additionally, an $m$-qubit quantum Fourier transform (QFT) involves at least one controlled rotation between any two qubits, and $\binom{m}{2}$ in total. Therefore,  a quantum device with all-to-all connectivity is preferred to limit the use of \textsf{SWAP} gates.

\begin{figure} 
    \centering
  \subfloat[Semiclassical QPE circuit\label{fig:semiclassical QPE}]{%
       \includegraphics[width=.5\linewidth]{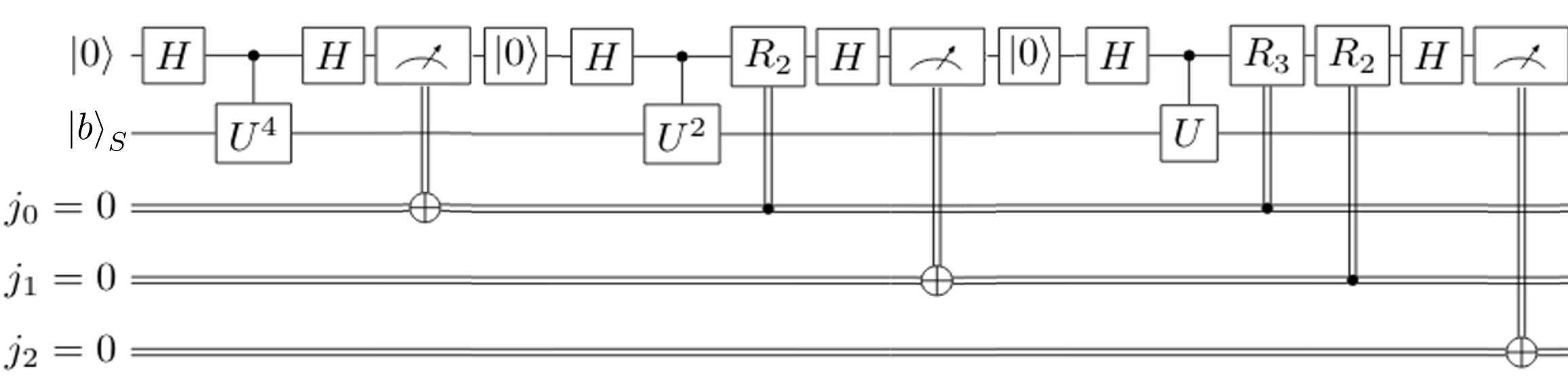}}
    \hfill
  \subfloat[Standard QPE circuit \label{diagram:qpe}]{%
        \includegraphics[width=.5\linewidth]{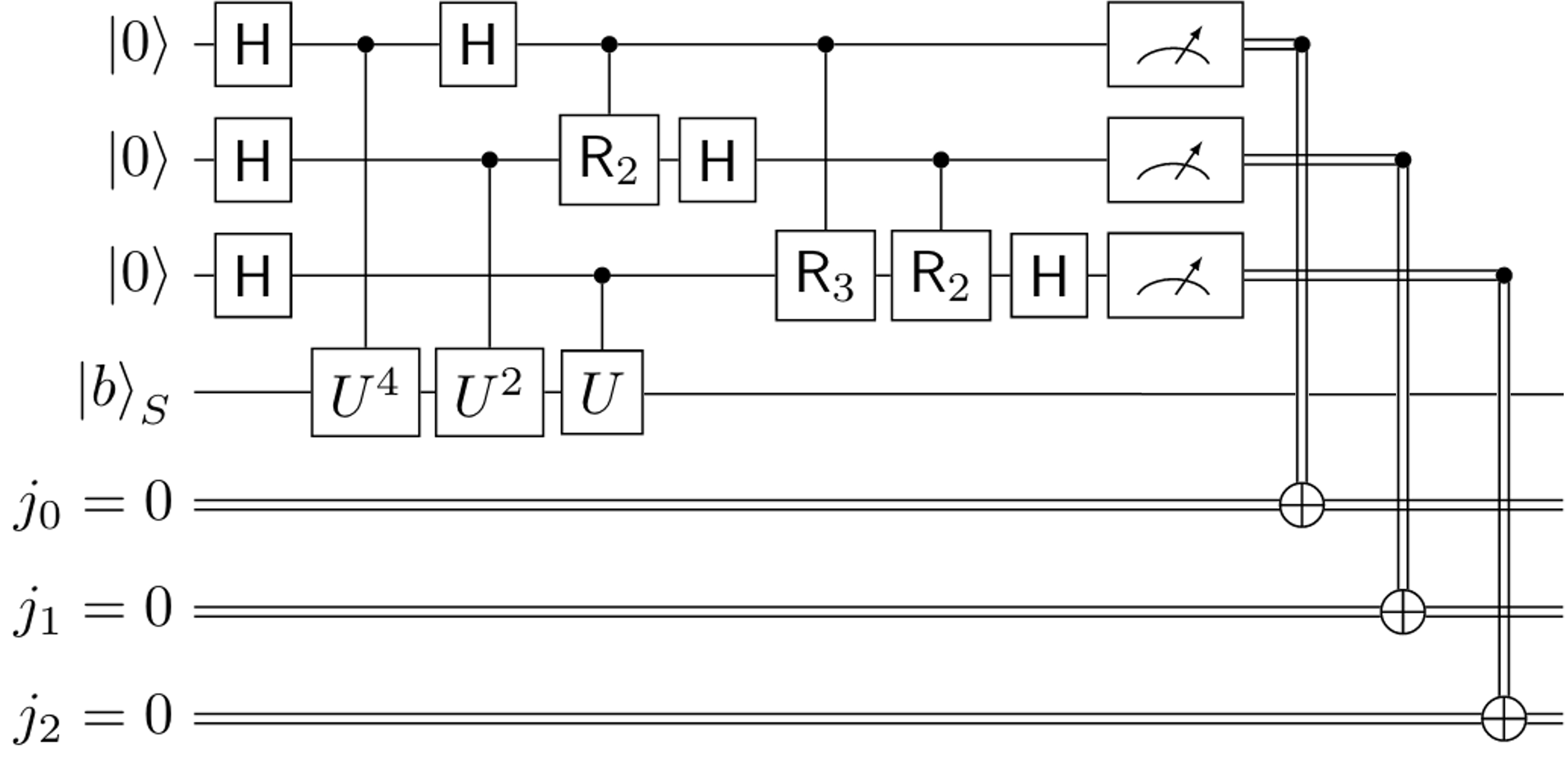}}
  \caption{Circuits for estimating the eigenvalues of the unitary operator $\mathbf{U}$ to three bits using semiclassical QPE (a) or QPE (b). $S$ is the register that $\mathbf{U}$ is applied to, and $j$ is a classical register. $\mathsf{H}$ refers to the Hadamard gate and $\mathsf{R}_\mathsf{k}$,  for $\mathsf{k}=2,3$, are the phase gates.}
  \label{fig1} 
\end{figure}

A phase estimation variant \cite{shors2nplus3, mosca1998hidden} has been identified that better lends itself to NISQ hardware and utilizes the semiclassical inverse  QFT\cite{semiclassicalqft}. We will refer to this variant as semiclassical QPE. This is a nonunitary version of the inverse QFT that estimates each bit of the eigenvalue sequentially. A diagram of this variant is displayed in Figure \ref{fig:semiclassical QPE}, which shows how to efficiently estimate the eigenvalues of the unitary operator $\mathbf{U}$ to three bits of precision. This variant leverages techniques for dynamic quantum circuits made available on recent hardware: mid-circuit measurements, ground-state resets, conditional logic, and qubit reuse.
This semiclassical variant of QPE has been previously demonstrated on superconducting quantum hardware \cite{C_rcoles_2021}. However, to the best of our knowledge, our work is the first one benchmarking its performance on trapped-ion devices, and the first to incorporate it into a quantum algorithm executed end-to-end on commercially available hardware for solving a practical problem.

The semiclassical QPE is mathematically equivalent to performing the original inverse QFT and measuring the eigenvalue register. 
It is also similar to iterative QPE (iQPE) \cite{Dobek2007iqpe} to the extent that it only requires one ancillary qubit to achieve arbitrary bit precision of the eigenvalues. A limitation of iQPE, however, is that it requires the initial state to be an eigenvector of $\mathbf{U}$ in order to estimate its corresponding eigenvalue. %This prior knowledge of the eigenvectors could be used to directly solve the QLSP. 
Conversely, semiclassical QPE can estimate eigenvalues without prior information of the eigenvectors. The description of how to implement the semiclassical QPE is shown below in Algorithm \ref{alg:conditional logic_qpe_algo}.

\begin{algorithm}[t]
\SetAlgoLined
\caption{Semiclassical Phase Estimation\cite{shors2nplus3, mosca1998hidden}}
\label{alg:conditional logic_qpe_algo}
%\begin{algorithmic}
Input: a bit precision $m$, unitary operator $U$, and input quantum state $\ket{b}$ \\
Step 1: Initialize one only ancilla qubit at $\ket{0}$, a register $S$ to the $\ket{b}$ state, and $m$ classical registers $\{j_0, j_1, \dots, j_{m-1}\}$ to store the results measured in the ancilla qubit.\\ 
Step 1: Apply Hadamard $\mathsf{H}$ to the ancillary qubit \\
 Step 2: Apply the unitary $\mathbf{U}^{2^{m-1}}$ on the $S$ register conditioned on the ancilla qubit. \\
Step 3: Apply Hadamard $\mathsf{H}$ to the ancillary qubit. \\ 
Step 4: Measure the ancilla qubit and store the results on the classical register $j_0$. \\
\For{$i=2 \cdots m$}{
Step 5.1: Apply Hadamard $\mathsf{H}$ to the ancillary qubit\\ 
Step 5.2: Apply the unitary $\mathbf{U}^{2^{m-i}}$ on the $S$ register conditioned on the ancilla qubit. \\
\For{$k=i\cdots 2$}{
Step 6.1: Apply the phase gate $R_{k}$ on the ancilla qubits conditioned on the classical register $j_{i-k}$ \\
}
Step 7: Apply Hadamard $\mathsf{H}$ to the ancillary qubit \\
Step 8: Measure the ancilla qubit and store the results on the classical register $j_{m-i}$ 
}

\end{algorithm}

The semiclassical QPE is used to perform the eigenvalue estimation in Hybrid HHL$^{++}$, and it has two properties that make it more suitable for near-term devices than the standard version of phase estimation:
\begin{enumerate}
    \item the procedure requires only one ancillary qubit for an arbitrary bit precision,
    \item and it replaces two-qubit gates with one-qubit gates controlled by classical registers.
\end{enumerate}
These differences can be observed by comparing Figures \ref{fig:semiclassical QPE} and \ref{diagram:qpe}.

Our work proposes to modify the eigenvalue estimation step (step 1) of Hybrid HHL to use the semiclassical QPE. However, in step 2, we still utilize the textbook implementation of QPE prior to the eigenvalue inversion step. The benefits of this will be highlighted shortly, where we discuss our heuristic for reducing the depth of the eigenvalue inversion circuit. In the \emph{Results} section, we will report data from extensive benchmarking of the semiclassical QPE on a trapped-ion device.

Note that the semiclassical phase estimation cannot be used for the eigenvalue inversion since it repeatedly collapses the quantum state of the ancillary qubit through mid-circuit measurements. Therefore, this eigenvalue estimation procedure cannot be incorporated as part of a deeper circuit that relies on the quantum states encoded in this ancillary register. Thus, we specifically utilize the semiclassical procedure for solely estimating the rotation angles, which can be done to arbitrary precision with a single ancillary qubit.

Recall that the eigenvalue inversion for Hybrid HHL consists of applying the following controlled rotation for each estimated eigenvalue $\lambda$:
\begin{align}
|2^m\gamma\lambda\rangle\langle2^m\gamma\lambda| \otimes \mathsf{R}_{\mathsf{Y}}(2\sin^{-1}(C/\lambda)) + [\mathbb{I} - |2^m\gamma\lambda\rangle\langle2^m\gamma\lambda|]  \otimes \mathbb{I}.
\end{align}
However, as observed by Lee \etal~, the purpose of QPE in the eigenvalue estimation and inversion steps is quite different. Specifically, in step (1) of Hybrid HHL the goal is to estimate the rotation angles $2\sin^{-1}(C/\lambda)$. This is where the most precision is required. In step (2), the goal is to just be able to just find a bijection between distinct eigenvalue estimates in step (1) and the values the control register of the rotation can take. More specifically, we just want

\begin{align}
\label{eqn:modified_inversion}
|f(\lambda)\rangle\langle f(\lambda)| \otimes \mathsf{R}_{\mathsf{Y}}(2\sin^{-1}(C/\lambda)) + [\mathbb{I} - |f(\lambda)\rangle\langle f(\lambda)|]  \otimes \mathbb{I},
\end{align}
where $f$ is bijection from the set of eigenvalues with $\lvert c_{\lambda}\rvert > 0$ and $k$-bit integers, \emph{where $k$ may be less than $m$}. This means in some cases, we can actually run a QPE that uses fewer qubits in step (2) of Hybrid HHL than step (1). The approach of Ref.~\cite{lee2019hybrid} utilized a concept called eigenmeans. This technique can result in $k <  m$ when the bit representations of $2^m\gamma\lambda$ agree on the first $k$ bits but all differ on the last $m-k$ bits. Unfortunately, this appears to be a strong assumption and the authors also showed it working for a single example. In addition, an important point to emphasize is that the eigenmean approach does not allow for reducing the number of ancillary qubits used by the phase estimation circuit. The use of eigenmeans only allows for reducing the number of controlled rotations and potentially the qubits they control on, not the size of the QPE circuit.

In this section, we present an alternative to the eigenmeans approach that is more generic and allows for reducing the complexity of the overall HHL circuit, i.e., an eigenvalue inversion circuit sandwiched between two standard QPEs (see Figure \ref{diagram:algorithm}). %In fact, we apply it to a few different examples in the \emph{Results} section, showing the reduction in the number of qubits and rotations.}
The first step of our procedure is to utilize the spectral scaling algorithm of Section \ref{sec:spectral_scaling} (Algorithm 1 and 2) to determine a scaling factor and a number of qubits $m$ that is sufficient to resolve the relevant eigenvalues of $\mathbf{A}$. The observed eigenvalues will be used to compute the rotation angles for Equation \eqref{eqn:modified_inversion}. As mentioned earlier, the number of ancillary qubits used by the semiclassical QPE does not increase with the desired precision. Thus, we will make the following simplifying assumption that there exists a $\gamma \in \mathbb{R}$ and integer $m$ such that for all eigenvalues $\lambda$ of $\mathbf{A}$: $2^m\gamma\lambda \in \mathcal{B}_{m}$. We will denotes the $m$-bits of $2^m\gamma\lambda$ by $\lambda^{(1)}\cdots\lambda^{(m)} = \lambda^{(1:m)}$. We will also assume that $2^m\gamma\lambda \neq 2^{m-1} -1$.

Suppose that instead we use an $k$-bit QPE for $k < m$,  which performs the following unitary \cite{nielsen00}:
\begin{align}
\sum_{\lambda}c_{\lambda}\ket{\psi_{\lambda}}\ket{\mathbf{0}_{m}} \mapsto \sum_{\lambda}c_{\lambda}\ket{\psi_{\lambda}}\otimes \sum_{\ell=0}^{2^m -1}2^{-k}\frac{1- e^{2\pi i2^{k}(2^{-m}\lambda^{(k+1:m)} - 2^{-k}\ell)}}{1- e^{2\pi i(2^{-m}\lambda^{(k+1:m)} - 2^{-k}\ell)}}\ket{\lambda^{(1:k)} + \ell \mod 2^{k}},
\end{align}
where by assumption, $\frac{\lambda^{(1:m)}}{2^m} \in [\frac{\lambda^{(1:k)}}{2^k}, \frac{\lambda^{(1:k)} + 1}{2^k}]$. Specifically, we want the smallest $k$ such that only one $\lambda^{(1:m)}$ lies in each of these intervals. This allows us to perfectly distinguish the eigenvalues of $\mathbf{A}$.
It is well known \cite{Brassard_2002} that the probability mass that phase estimation assigns to this region is,
\begin{align}
&2^{-k}\frac{1- e^{2\pi i2^{k}(2^{-m}\lambda^{(k+1:m)})}}{1- e^{2\pi i(2^{-m}\lambda^{(k+1:m)})}} + 2^{-k}\frac{1- e^{2\pi i2^{k}(2^{-m}\lambda^{(k+1:m)} - 2^{-k})}}{1- e^{2\pi i(2^{-m}\lambda^{(k+1:m)} - 2^{-k})}}\\ &\geq \frac{8}{\pi^2} \approx 0.81.
\end{align}
Thus if we were perform the following controlled rotation:
\begin{align}
(|\lambda^{(1:k)}\rangle\langle \lambda^{(1:k)}| + |\lambda^{(1:k)} +1\rangle\langle \lambda^{(1:k)} + 1|) \otimes \mathsf{R}_{\mathsf{Y}}(2\sin^{-1}(2^{m}C/\lambda^{(1:m)})) + \ket{\perp}\bra{\perp}  \otimes \mathbb{I},
\end{align}
and then invert QPE, we would obtain a state with at most a trace distance of $0.19$ from the state obtained by using an $m$-qubit (exact) QPE. Note that $\ket{\perp}\bra{\perp}$ corresponds to the projector onto the complement of $|\lambda^{(1:k)}\rangle\langle \lambda^{(1:k)}| + |\lambda^{(1:k)} +1\rangle\langle \lambda^{(1:k)} + 1|$. In addition, the trace distance can be reduced to $\epsilon$ by increasing the QPE precision by $\lceil \log\left (2 + \frac{2}{\epsilon}\right)\rceil$ additional qubits \cite{nielsen00}.

If there are two $\lambda^{(1:m)}$ that fall into adjacent intervals, then we can either increase $k$ to cause them to be further separated, or choose the angle of rotation controlled on their common endpoint to be an average of the reciprocals. In addition, it may be that we only need to control on $\lambda^{(1:k)}$, depending on where in the interval $\lambda^{(1:m)}$ falls. However, such a case could be an indication that we can further reduce the precision $k$.

One may notice that this procedure appears to work in a regime that is opposite to where the eigenmeans procedure introduced by Lee \etal~worked. Specifically, the above procedure saves when the higher-order bits of the eigenvalues differ, whereas, the eigenmeans procedure saves when the lower-order bits differ. Thus if $\exists$ $t \leq m$ s.t. $\lambda^{(1:t)}$ agree for all $\lambda$, i.e., the case of fixed eigenmeans coined by Ref.~\cite{lee2019hybrid}, then instead of only controlling on at most $2^{m-t}$ values from an $m$-qubit register, it actually suffices to perform QPE on $2^{t}\gamma\mathbf{A}$ with only $m-t$ qubits. The periodicity of the complex exponential with cause phase estimation to ignore the most significant bits. This reconciles the difference between our approach and the eigenmeans, and shows that our proposal works generically.

We summarize the overall procedure below.
\begin{enumerate}
    \item  Using semiclassical QPE, optimize the scaling factor $\gamma$ using the algorithms from the previous subsection. Determine a number of bits $m$ that is sufficient to resolve eigenvalues of $\gamma\mathbf{A}$ to high precision. These are the $m$-bit estimations $\lambda^{(1:m)}$. We can also estimate the probability of the estimates and only accept ones above a certain threshold (for exact estimation this corresponds to thresholding on $\lvert c_{\lambda} \rvert$). This is done in Step 1 and 2 from Figure \ref{diagram:algorithm}.

    \item Reduce the precision of QPE starting from $m$ to obtain the smallest $k \leq m$ s.t. $\frac{\lambda^{(1:m)}}{2^m} \in [\frac{\lambda^{(1:k)}}{2^k}, \frac{\lambda^{(1:k)} + 1}{2^k}]$ is satisfied (i.e., no two $m$-bit estimates lie between two $k$-bit estimates). Potentially a different $\gamma$ can be chosen as well. If all $\lambda^{(1:m)}$ agree on the first $t$-bits, multiply the matrix by $2^t$. If the intervals for two $\lambda^{(1:m)}$ share an endpoint, then we can assign the rotation angle to be the average of the reciprocals of the $\lambda^{(1:m)}$. This is done in Step 3 from Figure \ref{diagram:algorithm}.

    \item Construct the HHL circuit by performing a $k$-bit QPE. Then for each $\lambda^{(1:m)}$ apply the rotation:
    \begin{align}
        (|\lambda^{(1:k)}\rangle\langle \lambda^{(1:k)}| + |\lambda^{(1:k)} +1\rangle\langle \lambda^{(1:k)} + 1|) \otimes \mathsf{R}_{\mathsf{Y}}(2\sin^{-1}(2^{m}C/\lambda^{(1:m)})) + \ket{\perp}\bra{\perp}  \otimes \mathbb{I},
    \end{align}
    then invert the $k$-bit QPE. Depending on where the eigenvalue falls in the interval, controlling on $\lambda^{(1:k)}$ only may suffice. As mentioned earlier, this is an indication that we can potentially reduce the precision to cause  $\lambda^{(1:m)}$ to fall between to $k$-bit estimates. This is done in Step 4 from Figure \ref{diagram:algorithm}.
\end{enumerate}

The benefit of this procedure is that when running HHL, both the size of the standard QPE and the eigenvalue inversion circuit can be significantly reduced. This saves on both depth and ancillary qubits. The use of the semiclassical QPE allows for high-precision estimates with only one ancillary qubit. Thus, the whole procedure can be substantially more near-term friendly. However, we emphasize that this approach is more of a heuristic, yet, we believe it is well motivated by the discussion in this subsection. The main message is that once we have classical access to the $m$-bit eigenvalue estimates, the only goal of the  phase estimation in the HHL circuit is to produce a mapping that separates distinct eigenvalues.

\section{Results}
\label{sec:results}

In this section, we detail the outcomes from the experiments with Hybrid HHL$^{++}$ on the Quantinuum H-series devices. We also include results from a noisy emulator of the machine provided by Quantinuum and noiseless simulation. These demonstrations are at a larger scale than prior hybridizations of HHL and showcase both the benefits of our techniques and the advancement of quantum hardware. We believe that this implies that  Hybrid HHL$^{++}$ is a good application-oriented benchmark for small hardware.

Our application of choice is portfolio optimization formulated as a linear system, hence the emphasis of an application-oriented benchmark. This is based on the formulation of Rebentrost and Lloyd, which we detail below.

\subsection{Portfolio Optimization as a Linear System}

HHL has been proposed as a possible solver for a specific portfolio-management problem~\cite{rebentrost2018quantum}, known as mean-variance portfolio optimization.
Given a set of $\Tilde{N}$ assets,
this problem requires the following quantities as inputs: the \textit{historical covariance matrix} $\Sigma \in \mathbb{R}^{\tilde{N} \times \tilde{N}}$, the \textit{expected returns} $\mathbf{r} \in \mathbb{R}^{\tilde{N}}$, and the \textit{prices} $\mathbf{p} \in \mathbb{R}^{\tilde{N}}$ of the  assets. Its objective is to minimize the \textit{risk}, represented by the quadratic form $\mathbf{w}^\mathsf{T} \Sigma \mathbf{w}$, subject to the desired \textit{expected total return} $\mu \in \mathbb{R}$ and \textit{budget} 
$\xi \in \mathbb{R}$. The solution $\mathbf{w} \in \mathbb{R}^{\tilde{N}}$ is the \textit{allocation vector} that weighs each asset in the portfolio. Since the solution is a weight vector, the budget is only a scaling parameter and can be set to $1$. Moreover, in order to compare portfolio performances, the return of a portfolio is usually expressed as a percentage instead of a monetary amount. With a simple change of variables, the problem can be reformulated in these terms.

This problem can be stated as a convex quadratic program: 

$$
\label{opt_equation}
\underset{\mathbf{w} \in \mathbb{R}^{\tilde{N}}}{\text{minimize}}~ \mathbf{w}^\mathsf{T} \Sigma \mathbf{w}~:~\xi = \mathbf{p}^\mathsf{T} \mathbf{w},~\mu = \mathbf{r}^\mathsf{T} \mathbf{w}
$$

This quadratic program can be reformulated as a linear system by using the method of Lagrange multipliers, resulting in the following equation:

\begin{equation}
    \centering
    \label{linear_system}
    \begin{bmatrix}
    0 & 0 & \vec{r}^\mathsf{T}  \\
    0 & 0 & \mathbf{p}^\mathsf{T} \\
    \mathbf{r} & \mathbf{p} & \Sigma
    \end{bmatrix}
    \begin{bmatrix}
    \eta \\ \theta \\ \mathbf{w}
    \end{bmatrix}
    =
    \begin{bmatrix}
    \mu \\ \xi \\ \mathbf{0}
    \end{bmatrix}
\end{equation}
where $\eta, \theta \in \mathbb{R}$ are the Lagrange multipliers.
We will denote this linear system by $A \mathbf{x} = \vec{b}$, with $A \in \mathbb{R}^{N \times N}$ and $\mathbf{x}, \mathbf{b} \in \mathbb{R}^{N}$, where $N=\tilde{N} + 2$. A quantum state representing the solution, up to a normalization constant, can be obtained by solving the corresponding QLSP using HHL. This can be done because the covariance matrix, $\Sigma$, is Hermitian, and so $A$ is Hermitian too. The resulting quantum state $\ket{\mathbf{x}} = \ket{\eta, \theta, \mathbf{w}}$ allows us to recover $\ket{\mathbf{w}}$. 

Following this approach, Rebentrost and Lloyd \cite{rebentrost2018quantum} have shown how to use the quantum state produced by HHL to make calculations that are of interest to the financial industry. For instance, given the optimal portfolio state, one can measure the portfolio's risk, or compare, through a controlled-\textsf{SWAP} test \cite{patentBeusoleil}, the optimal portfolio to another candidate portfolio (e.g., one offered by a third party) that has been loaded onto a quantum state. The result of this comparison can then be used to decide which portfolio to invest in.

\subsection{Experiments on a Trapped-ion Device}
%In the following subsections we will benchmark our techniques for improving the performance of Hybrid HHL on near-term quantum hardware. Our aim is to execute and validate our approach on real quantum hardware supporting mid-circuit measurement, qubit reset and reuse, and quantum conditional logic. 

%\textcolr{red}{DH: i think we can remove this discussion and just direct the reader to the spec sheet.}
%All the experimental results that are presented in this article were obtained on trapped-ion Quantinuum H$1$ Systems given its support for mid-circuit measurements, qubit resets and reuses, and conditional logic \cite{Pino_2021}. This quantum processor uses a quantum charge-couple device architecture with three parallel gate zones in a linear trap. The quantum states are stored in the hyperfine states of twenty \textsuperscript{171}Yb\textsuperscript{+} atoms. All-to-all connectivity is implemented by rearranging of the physical location of qubits, which introduces a negligible amount of error. Typical single-qubit gate infidelity is $5\times 10^{-5}$ and typical two-qubit gate infidelity is $3\times 10^{-3}$. Typical error rate of state preparation and measurement is $3\times 10^{-3}$. Memory error per qubit at average depth-1 circuit (``idle error'') is $4\times 10^{-4}$. Additional details are available in \cite{h11datasheet}. \rev{For comparison, we also run the quantum algorithms on an emulator provided by Quantinuum \cite{h1edatasheet},
%which approximates the noise of H$1$ device, and in noiseless simulation}

In order to run the circuits on Quantinuum hardware, we transpiled and optimized the circuits from IBM's Qiskit \cite{aleksandrowicz2019qiskit} to the H-series devices' native gates using 
Quantinuum's pytket package \cite{Sivarajah_2020}.
As the error rates of two-qubit gates are an order of magnitude larger than those of one-qubit gates \cite{Pino_2021}, and the numbers of both gate types
are similar in the circuits used, we will only present the number of \textsf{ZZPhase} gates, the H-series native two-qubit gate, for the circuits. The \textsf{ZZPhase} is equivalent to $\mathsf{R}_\mathsf{ZZ}(\theta)$, and, up to one-qubit gates, it is realized via the M{\o}lmer-S{\o}rensen interaction \cite{Pino_2021}. The error is related to the magnitude of the angle of rotation $\theta$.

The data for the linear system matrix $\mathbf{A}$ comes from S\&P 500 stock data. Specifically, we utilized 2 assets, leading to a $4 \times 4$ matrix. The matrix is constructed using the procedure for formulating constrained portfolio optimization as a linear system that was highlighted in the previous subsection. We created six different problem instances and show results using these instances in noiseless simulation using the Qiskit QASM simulator and on the Quantinuum  trapped-ion hardware and noisy emulators. We took advantage of the emulator by executing the circuits with different numbers of shots to identify the minimum such that there are no significant changes in the results. Note that typically trapped-ion devices require significantly more time than superconducting devices to run perform a single shot. For reference, for the Quantinuum H-series devices, each shot of a QAOA with $p=2$ and 32 qubits takes 1 second \cite{moses2023race}. Thus, we also need to make a compromise on the number of shots to ensure that the experiments can be executed within a reasonable time frame.

The experiments are split into three categories resulting in different subsections. First, we show how the spectral scaling algorithm, introduced in Section \ref{sec:spectral_scaling}, works to determine the optimal $\gamma$ named $\gamma_{op}$. Then, with this optimal value of the scaling parameter, we perform a comparative analysis between the semiclassical and textbook phase estimation procedures on hardware in Section \ref{subsection:semiclassical QPE-benchmark}.  Lastly, we utilize the controlled-\text{SWAP} test on hardware to check the quality of the portfolio proposed by the algorithm in Section \ref{subsection:portfolio_hardware}. 

%Second, in Section \ref{subsection:compressing} we show experimental results of the calibration of the eigenvalue inversion circuit by using our procedure for reducing its complexity introduced in Section \ref{section:eigenvalue_inversion}.

% \subsection{semiclassical QPE for Eigenvalue estimation}
% \label{section:semi_classical_qpe}

% %%%%
% \subsubsection*{Performance on Hardware}
% \label{subsection:benchmark_conditional logic_qpe}

% We benchmarked the performance of both the standard QPE and semiclassical QPE for estimating the eigenvalues of the operator $U := e^{i A 2\pi \gamma}$ of a $4 \times 4$ matrix $A$ scaled by a parameter $\gamma$ and applied to an initial state $\ket{b}$. A technique for selecting the scaling parameter will be presented Section \ref{sec:evol_time_optimization}. %using the initial state $\ket{b}$ and $\gamma = 100$. 
% The matrix and initial state used correspond to  a constrained portfolio-optimization problem with two \mbox{S\&P 500} assets. The way portfolio optimization problem is cast as a QLSP is explained in Section \ref{subsec:PortfolioOptimizationAsLinearSystem}.% The matrix $A$ of the linear system is of size $4\times4$. 

\subsubsection{Spectral Scaling Algorithm Benchmarks} \label{subsection:Estimating spectrum with QPE}
%\textcolor{red}{Only discuss how scaling parameter helps to better resolve eigenvalues and waste less of the allotted precision (for semiclassical QPE this corresponds to a depth reduction)}\\

In this section we present results obtained on hardware that show how Algorithm \ref{alg:time_evolution_optimization} and Algorithm \ref{alg:inital_guess_checker} optimize the scaling parameter $\gamma$, allowing for effectively estimating the relevant eigenvalues in the output distribution of quantum phase estimation. These proposed novel algorithms allow for reducing the resources needed to increase the bit precision to accurately identify the relevant eigenvalues. The optimized scaling parameter $\gamma_{op}$ makes the best use of the bit precision available. %to accurately distinguish the eigenvalues in the output distribution of quantum phase estimation. 
In this work, we leverage the semiclassical QPE, which, as explained before, lends itself to near-term hardware. What is more, the estimation of the eigenvalues with lower bit precision, thanks to using an optimal scaling parameter, also allows to reduce the number of qubits in the HHL circuit. %QLSP corresponding to the portfolio-optimization problem of two assets used for experimental evaluations in Section \ref{subsection:benchmark_conditional logic_qpe}.

\begin{figure}[h!]
    \centering
    \includegraphics[width=.5\linewidth]{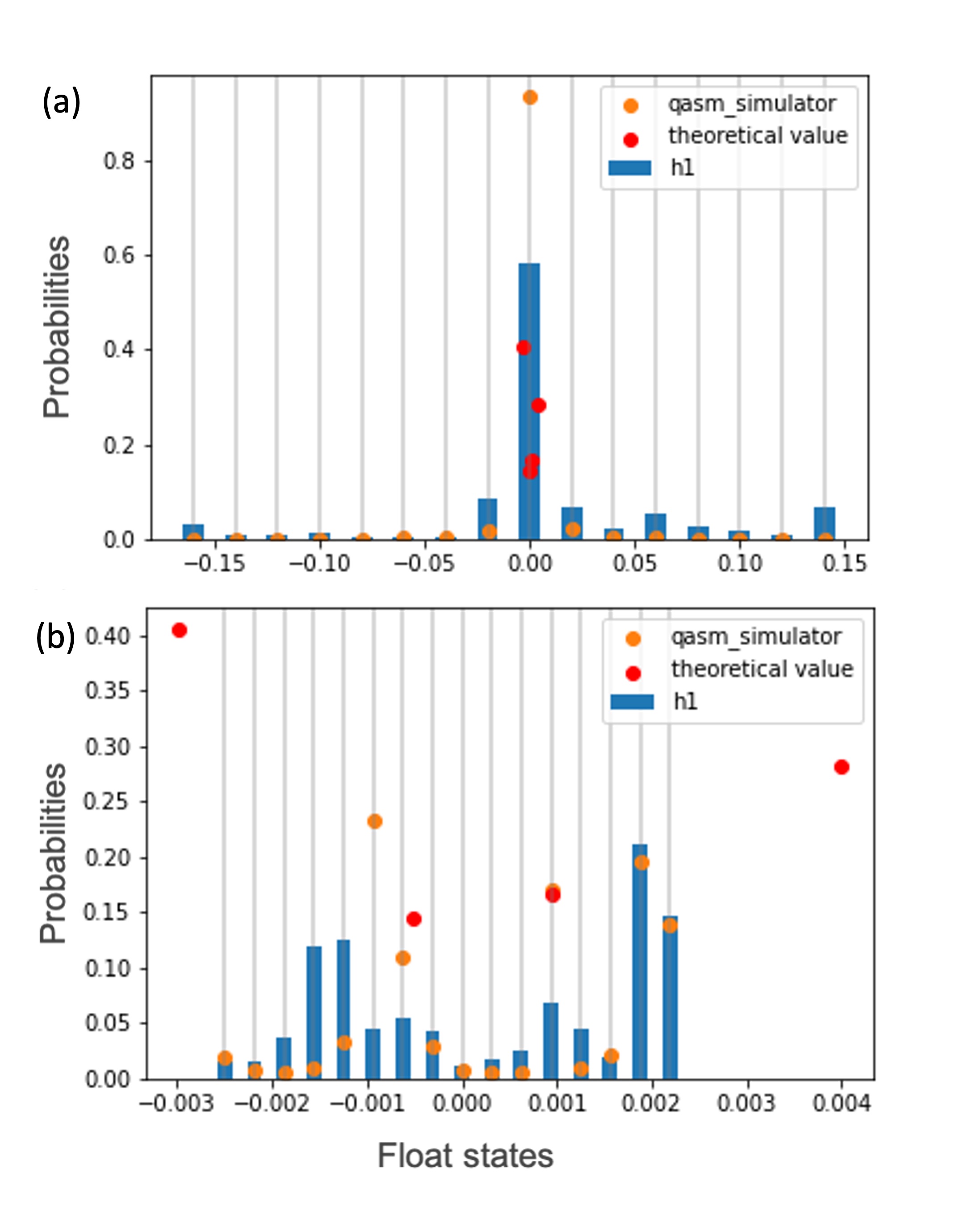}
\caption{Probability distributions over the four-bit eigenvalue estimates from the semiclassical QPE executions using $e^{i \mathbf{A} 2\pi \Gamma}$ for \mbox{$\Gamma = 50 \cdot 2^{-4}$} (a) and $\Gamma = 200$ (b). The blue bars represent the experimental results on the trapped-ion device with $2000$ shots. The theoretical (classically calculated) eigenvalues are represented with the red dots and the results from the noiseless simulator, Qiskit QASM simulator, are represented with the orange dots. As shown on (b), as the theoretical eigenvalues exceed the values we can encode, the distribution we observed shows that values overflowed.} 
\label{fig:Algorithms-two-output}
\end{figure}

To show how the optimization of $\gamma$ works, we select one of the problem instances corresponding to the optimization of a portfolio of two assets ($4 \times 4$ system) and we executed the semiclassical QPE one of the Quantinuum machines. We started with an initial guess of $0.02$ for $\lambda_\textup{max}$, corresponding to $\gamma=50$. We verified with \mbox{Algorithm \ref{alg:inital_guess_checker}} that $\alpha=0.02$ was indeed an overestimation of $\lambda_\textup{max}$, as required by \mbox{Algorithm \ref{alg:time_evolution_optimization}}. Following \mbox{Algorithm \ref{alg:inital_guess_checker}}, in the case of negative eigenvalues, we tested the validity of the initial guess $\gamma = 50$ with precision $n=4$ by running semiclassical QPE with \mbox{$\Gamma = \gamma \cdot 2^{-4}  = 50 \cdot 2^{-4}$}. The output probability distribution is shown in \mbox{Figure~\ref{fig:Algorithms-two-output} (a)}, together with the distribution obtained in noiseless simulation (``qasm\_simulator'') and the theoretical values of the eigenvalues, calculated analytically (``theoretical value''). We observe that the outputs from both the quantum device and noiseless simulation are  concentrated around zero, thus $\gamma = 50$ is a valid guess. An example for an invalid input would be $\gamma=3200$. The output distribution of semiclassical QPE with \mbox{$\Gamma= \gamma \cdot 2^{-4}  = 200$} is plotted on \mbox{Figure~\ref{fig:Algorithms-two-output} (b)}. As this distribution is not concentrated around zero, $\gamma=1600$ is not a valid guess. 

\begin{figure}[h!]
    \centering
    \includegraphics[width=.5\linewidth]{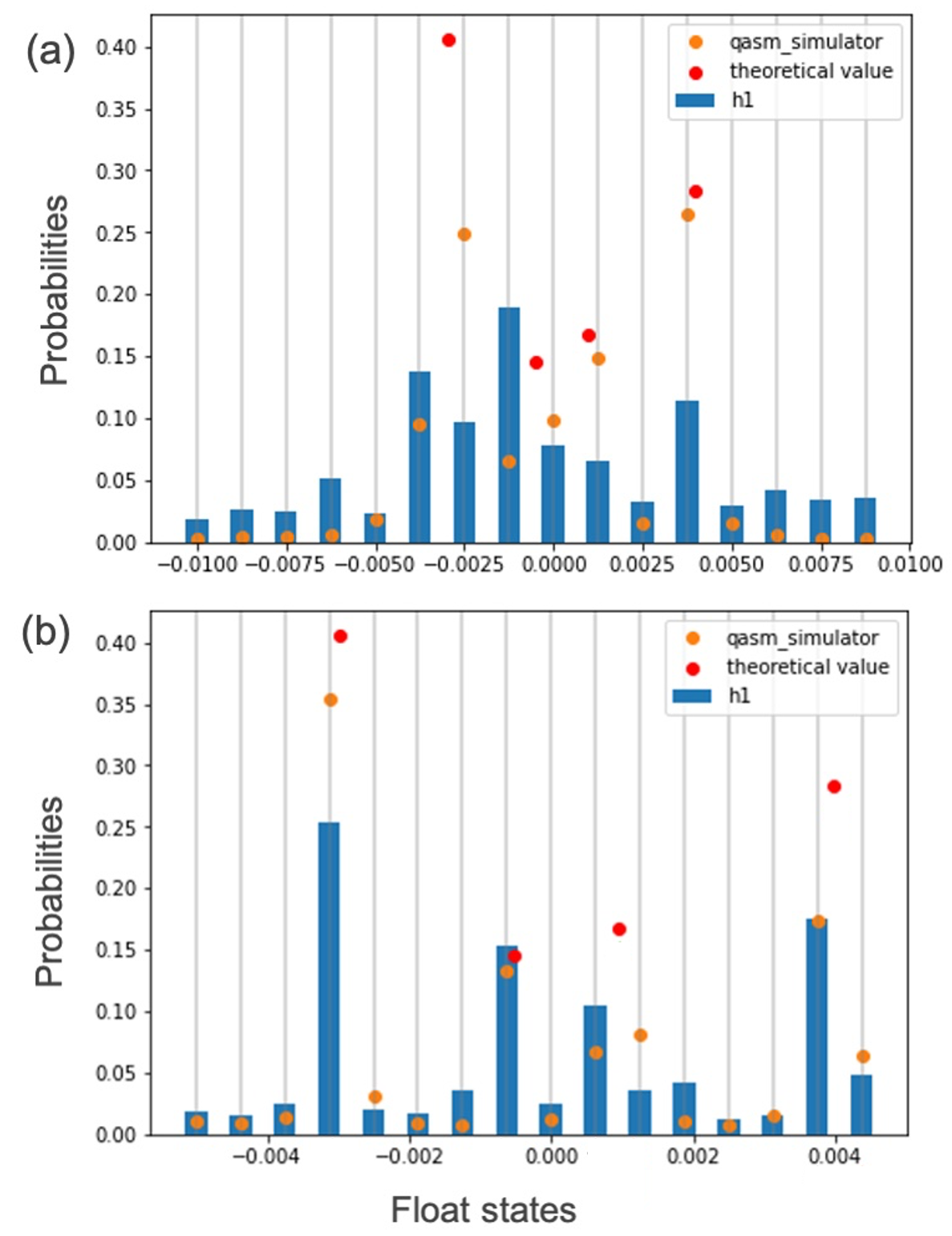}
\caption{Probability distributions over the four-bit eigenvalue estimates from the semiclassical QPE run using $e^{i A 2\pi \gamma}$ for $\gamma = 50$ (a) and $\gamma = 100$.The blue bars represent the experimental results on the trapped-ion device - $2000$ shots (a), $1000$ shots (b). The theoretical (classically calculated) eigenvalues are represented with the red dots and the results from noiseless simulation with the Qiskit QASM simulator are represented with the orange dots.} 
\label{fig:time_evolution_exp}
\end{figure}

Now that we confirmed $\gamma=50$ is a valid guess, we can execute Algorithm \ref{alg:time_evolution_optimization}. We ran semiclassical QPE for estimating the eigenvalues of $\mathbf{A}$ with this value of $\gamma$, and we used $\ket{\mathbf{b}}$ as the initial state. The output probability distribution obtained both on hardware and in noiseless simulation are displayed in Figure~\ref{fig:time_evolution_exp} (a), together with the theoretical values. The $x$-axis is binned into $16$ values, which are all of the possible four-bit estimates in decimal. They are represented by the grey, vertical lines. In the experiment, we only observed significant probabilities (blue bars) for states within the range $[-0.005, 0.005]$. 

Thus, in order to better distinguish the eigenvalues, we decreased the distance between bins by increasing the scaling factor $\gamma$. We did so by using our spectral scaling parameter optimization algorithm, \mbox{Algorithm \ref{alg:time_evolution_optimization}}, which increased $\gamma$ to $100$.
As explained before, we overestimated $\lambda_\textup{max}$ and hence underestimated $\gamma$ to avoid overflow.

We can see in Figure~\ref{fig:time_evolution_exp} (b) that using $\gamma=100$ makes the $x$-axis range smaller than in (a), while keeping the same number of bins. As a consequence, the bin intervals are smaller, and there is a better agreement between the theoretical eigenvalues, which are classically calculated, and the experimental probabilities. Moreover, we can see close concordance between the experimental and the simulation results (orange dots). The ability to better distinguish eigenvalues in (b) over (a) shows the importance of the scaling factr optimization procedure. For clarification, the distribution of classically calculated, theoretical, values in the plots are distributed according to $\{|c_{\lambda_i}|^2\}_{i=0}^{N-1}$, using the notation introduced in Section \ref{sec:rev_of_hybrid}.

\subsubsection{Semiclassical and Standard Quantum Phase Estimation Benchmarks \label{subsection:semiclassical QPE-benchmark}} 

The aim of this subsection is to present the results from benchmarking the semiclassical quantum phase estimation to estimate the eigenvalues on hardware. First, we compare its performance to estimate the eigenvalues up to different bit precisions. For the same problem instance we found the optimal $\gamma_{op}$ by analyzing results on hardware in previous subsection, we executed the semiclassical QPE to the unitary with this optimal $\gamma_{op}$ on the hardware to estimate the eigenvalues of $\mathbf{A}$ to different precisions: three, four and five. We compare the number of gates and qubits required for both the standard QPE and the semiclassical implementations in Table~\ref{table:2Qgates-standard-vs-conditional logic}. We can see that this variant of QPE employs fewer qubits and gates. As the precision grows, the number of two-qubit gates increases for both implementations. However, the number of two-qubit gates saved using this variant of QPE, instead of the standard implementation, grows quadratically as $n(n-1)$, with $n$ being the bit precision. Moreover, even though the precision in bits increases, the number of qubits in this implementation does not change. This contrasts with the linear growth in the standard QPE.

\begin{table}[h!]
\normalsize
 \centering
  \begin{tabular}{l|*{4}{c}}
               &        & $3$-bit & $4$-bit & $5$-bit \\ \hline
      Standard QPE & Gates  & 63      & 88      & 115 \\ 
               & Qubits & 5       & 6       & 7  \\ \hline
      Semiclassical QPE      & Gates  & 57      & 76      & 95 \\
               & Qubits & 3       & 3       & 3
  \end{tabular}
  \caption{Comparison of the number of two-qubit \textsf{ZZPhase} gates and qubits in both QPE implementations for estimating eigenvalues to different precisions.}
  \label{table:2Qgates-standard-vs-conditional logic}
  \end{table}

In order to quantify the performance of both implementations, we compared the empirical distributions of measurement results from the circuit execution on quantum hardware to the distribution obtained from noiseless simulation. One way to compare two probability mass functions, $p$ and $q$, is to use the fidelity metric \cite{nielsen00}: \begin{align}
F(p, q) = (\sum\limits_i \sqrt{p_i q_i})^2,
\end{align} where $F(p,q) \in [0, 1]$. This is a measure of the closeness of the measured quantum state, in the computational basis, on the quantum hardware to the quantum state measure on the noisy simulator. The higher this number, the better, as it means that the quantum state prepared on quantum hardware is closer to the one obtained from noiseless simulation. We approximate the variability we would expect if we were to repeatedly sample from the two unknown distributions and calculate the fidelity of the sample each time. For this, we use the bootstrap method that repeatedly resamples values from the original sample with replacement and calculates the fidelity, which is the statistic, of each resample. By doing this, we obtain the “bootstrap distribution” of the statistic. We report the median and the standard deviation of these distributions obtained with $9,999$ resamples.

We compare the achieved fidelity in both the standard and semiclassical phase estimation for three  different bit precisions, in Table \ref{table:fidelity-standard-vs-conditional logic}. It can be seen that the computed fidelity metrics for the two implementations are similar for three-bit estimations. Here the number of saved two-qubit gates using semiclassical QPE is small. In addition, reducing the number of qubits does not overcome potential errors due to mid-circuit measurements and resets.  

\begin{table}[h!]
\normalsize
 \centering
  \begin{tabular}{l|*{3}{c}}
     & $3$-bit & $4$-bit & $5$-bit \\ \hline
      Standard QPE & 98.6$\pm$1.2 & 90.4$\pm$1.7 & 42.6$\pm$2.4 \\
      Semiclassical QPE  & 98.1$\pm$1.4 & 95.0$\pm$1.6 & 43.2$\pm$1.0 \\
  \end{tabular}
  \caption{Fidelity expressed in \% between the probability distributions from the QPE experiments ran on quantum hardware and noiseless simulation, with $2000$ shots each. We report the median and the standard deviation of the bootstrap distributions. We utilized $2000$ shots, since we observed on the noisy emulator that a higher number does not change the fidelity significantly.}
  \label{table:fidelity-standard-vs-conditional logic}
\end{table}

When we increase the precision to four and five bits, the circuits in both implementations deepen, and therefore, we see a drop in fidelity. Nevertheless, as shown in Table~\ref{table:2Qgates-standard-vs-conditional logic}, conditional logic, mid-circuit measurement, and qubit reset and reuse, result in the semiclassical QPE circuit being shallower than the standard implementation. As a consequence, the achieved fidelity with semiclassical QPE is, in median, still higher than the standard QPE. In both implementations, the decay of the  fidelity for five-bit precision can be explained by the number of gates approaching the limit supported by current devices.

Another benchmark we perform is the following. Given a fixed bit precision of four bits, we repeat the procedure for the remaining five problem instances. For this, we followed the steps in Algorithm \ref{alg:time_evolution_optimization} and \ref{alg:inital_guess_checker}, and we performed the semiclassical QPE on the Quantinuum's emulator.. With this, the optimized scaling parameter $\gamma_{op}$ was obtained for each problem instance. We used the obtained value of  $\gamma_{op}$ to perform the semiclassical QPE  applied to the unitary $e^{2\pi i\gamma_{op}\mathbf{A}}$ on hardware. The  eigenvalues were estimated up to four bits. We quantify the performance of this variant of QPE by calculating the fidelity between its output distribution obtained both on the emulator and on hardware with the one obtained with noiseless simulation. The total number of qubits of the circuits is three and the median and standard deviation of two-qubit \textsf{ZZPhase} depth across the circuits is $63.00 \pm 4.60$.
%and the depth is $137.0 \pm 9.5$.
% together wit the characteristic of the constructed HHL circuit controlling on the set of eigenvalue estimates obtained through a detailed post-processing of the outputdistribution of the semiclassical QPE circuits corresponding to each of the problem instances. 

%--------semiclassical QPE---------
%depth: 62.0 1.0
%qubits: 3.0 0.0
%--------HHL---------
%depth: 263.5 1.5
%qubits: 10.0 0.0
%max depth 265

\begin{figure}[h!]
    \centering
    \includegraphics[width=.5\linewidth]{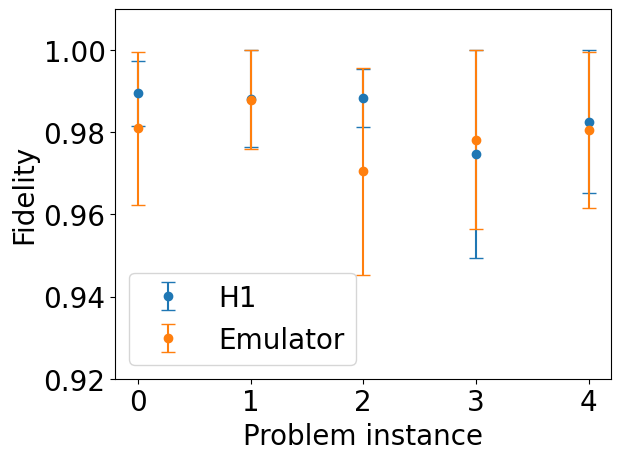}
\caption{Fidelity between the output distribution of the semiclassical QPE executed on both the  hardware and its emulator with $400$ shots and the output distribution obtained in noiseless simulation, respectively, as a function of the portfolio-optimization problem instances. The semiclassical QPE is performed to the unitary corresponding to the optimal scaling parameter obtained with the proposed algorithms using the output of the circuits on the emulator. We utilized this number of shots as it has been seen on the noisy emulator of the hardware that for a higher number, there are no significant changes. It is displayed the median and standard deviation, as error bars, of the bootstrap distributions. } 
\label{fig:conditional logic_qpe_fidelity_plot}
\end{figure}

In Figure \ref{fig:conditional logic_qpe_fidelity_plot} we compare the fidelity between the output distribution from both hardware and the emulator  with the distribution obtained in noiseless simulation, respectively. We report the results for five different problem instances and display the median and the standard deviation of the bootstrap distributions corresponding to the fidelity. We first observe that the obtained fidelity on the quantum device is very high. This high fidelity allows us to utilize the output distribution obtained with this variant of quantum phase estimation to accurately estimate the eigenvalues. Additionally, we observe that the results on hardware are on average at least as good as results from its emulator, showing that the emulator gives a good lower bound of what can be expected on hardware.

\subsubsection{Validating Portfolio Quality on Hardware \label{subsection:portfolio_hardware}}

Now that we have benchmarked all the components required to perform Hybrid HHL$^{++}$, in this section, we check the quality of  the solution produced by the algorithm for the constrained portfolio optimization problem with two assets. We followed the steps presented  for  Hybrid HHL$^{++}$ to find the optimal scaling parameter $\gamma_{op}$ for each problem instance, and we estimated the set of eigenvalues up to four bits with the semiclassical QPE executed on the emulator. Then, the eigenvalue inversion component was constructed using the heuristic method introduced in Section \ref{section:eigenvalue_inversion}. We found the proper $k$ for each problem instance such that $k \leq 4$ and each estimate is in separate bin in the output distribution of the semiclassical QPE utilizing the optimal scaling parameter. Note that $k$ is the number of ancillary qubits utilized and the angles of the rotations are calculated with the higher resolution $m$-bit estimates. Note that the $k$ depends on each problem instance as well as the size of the set, which also depends on the classical post-processing made over the probability distribution. The compression heuristic helps when the number of qubits, $k$, used by the standard QPE in the HHL circuit is smaller than the bit precision, $m$, used by the semiclassical QPE to estimate the eigenvalues. In particular for the problems considered, the set size varied between four and six. %\rev{This implementation of HHL is reduced because, as explained in the previous sections, the efficient finding of the optimal scaling parameter allows for reducing the number of ancilla qubits required in the HHL circuit. Moreover, the conditioning of the rotations on the set of eigenvalue estimates also allows to reduce the circuit depth because we can control the rotations on a smaller set in comparison to controlling on all possible binary state as a results of lacking of information about the eigenvalues. Note that lacking of this information and controlling on all the possible four-bit string would require to control on a set of size of, in the worse case, $2^k \leq  2^4 = 16$, dramatically increasing the circuit depth}. 

A way of quantifying how good the HHL algorithm is at solving the portfolio optimization problem is by comparing the estimation of the best portfolio to the classical calculated solution is by using a controlled-\textsf{SWAP} test \cite{patentBeusoleil} between the quantum states that represent them. The controlled-\textsf{SWAP} test can be used to compute the magnitude of the inner product between the quantum state that represents the allocation vector produced by HHL and the classical solution loaded onto a quantum state \cite{buhrman2001quantum}.

We added a new qubit to the HHL circuit called the \textit{swap ancillary qubit}, and we loaded the normalized classically calculated solution to the linear system, i.e., $\mathbf{x}_c$, onto this qubit. The circuits also have one rotation ancillary qubit. We ran HHL and the controlled-\textsf{SWAP} test between the HHL output state $\ket{\mathbf{x}}$ and the quantum state encoding the classical solution $\ket{\mathbf{x}_c}$ on quantum hardware for the five problem instances of portfolio optimization considered. %The number of qubits depend on each instance and the total is $10$ qubits. The median and the standard deviation of the \textsf{ZZPhase} depth across the different circuits is 263.5$\pm$ 1.5 and the depth is $498.5\pm28.5$. 

%We executed the circuits in the Qiskit statevector simulator and on the Quantinuum H$1$ hardware for the three different approaches discussed before: the uniformly controlled rotation gate and the proposed eigenvalue inversion circuit that inverts on the estimates of the relevant eigenvalues, conditioning on four and six estimates respectively. These are the sets introduced in Section \ref{subsection:Estimating spectrum with QPE}. 

We measured the rotation and swap ancillary qubits. With the results from the measurements of these two qubits using $1000$ shots, we calculated the inner product between the quantum states as: \begin{align}
\sqrt{2 \frac{P(\ket{10}) }{P(\ket{10}) + P(\ket{11})} - 1},
\end{align}
where $P$ represents the probability of measuring the respective quantum state. The most-significant qubit corresponds to the rotation ancilla, where a post-measurement state of $\ket{1}$ corresponds to success. A uniform distribution would result in a inner product equal to 0.

\begin{figure}[h!]
    \centering
    \includegraphics[width=.5\linewidth]{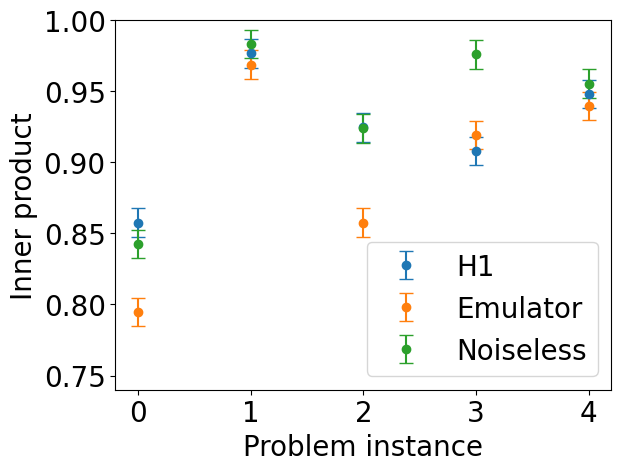}
    %\subfloat[\centering]{{\includegraphics[width= 7.8 cm]%{pictures/inner_product_hardware.png} }}
%    \qquad
%    \subfloat[\centering]{{\includegraphics[width=7.8 cm]{pictures/relative_inner_product_hardware.png} }}%
\caption{Inner product between the output quantum state of the HHL circuit and the analytical solution loaded as a quantum state for each of the portfolio-optimization problem instances with two assets. This metric quantifies the quality of the solution obtained. It is displayed the inner product corresponding to the execution of the circuits on  hardware, its emulator with approximated noise and in noiseless simulation with $1000$ shots. We utilized this number of shots as it has been seen on the noisy emulator of the hardware that for a higher number, there are no significant changes. The error bars correspond to the sampling error $\epsilon$ corresponding to $\mathcal{O}(\nicefrac{1}{\epsilon^2})$ shots. }
\label{fig:hhl_inner_prod}
\end{figure}

We compare the inner product obtained on the Quantinuum H$1$ hardware to the results obtained on its emulator with approximated noise and in noiseless simulation. Figure \ref{fig:hhl_inner_prod} shows the inner products corresponding to five portfolio-optimization problems considered corresponding to two assets. We first observe that the inner products obtained on hardware are very high, indicating that the solution obtained (the optimal portfolio) is very close to the analytical solution. We see that the results obtained on hardware either overlap with the results obtained in noiseless simulation or they are worse, as it occurs with one problem instance. Note that the error corresponds to the sampling error $\epsilon$ due to the $\mathcal{O}(\nicefrac{1}{\epsilon^2})$ shots. The overlapping or the closeness of these values show how good the reduced HHL circuits perform on hardware. Additionally, we note that the results on hardware are on average at least as good as results from its emulator, as expected.

We summarize the results obtained across the instances in Table \ref{table:inner_product_harware}, where we compare the median and the standard deviation of the inner products obtained across the problem instances, together with the circuits' characteristics. We note that the results obtained on real hardware are, in median, very close to one and although it is worse than the ones in noiseless simulator, this difference is only of $4.17 \%$ in median. 

These high inner product values obtained on hardware show how the proposed Hybrid HHL$^{++}$ allows for solving on real quantum hardware, with high accuracy, linear system problems corresponding to portfolio optimization with two assets. Our novel contributions to improve upon the Hybrid HHL algorithm has reduced the complexity of HHL circuit. By doing this, we have managed to execute the algorithm on hardware and solve, with high accuracy, a practically-relevant problem. 

%\rev{Figure \ref{fig:hhl_inner_prod} (b) displays the ``relative inner product'', which is relative to the values obtained by executing the circuits on noiseless simulation. This metric quantifies the performance of the quantum hardware in comparison to noiseless simulation. We observe that the results obtained on hardware are very close to the ones in noiseless simulation and actually, for some instances, the output is even better. This is due to noise in the hardware actually contributing to obtain a higher inner product. As expected, we additionally observe that the results on emulator are at least as good as the ones on hardware.}

\begin{table}[h!]
\normalsize
 \centering
  \begin{tabular}{l|l|l|l|l}
            &                         & \multicolumn{3}{c}{Inner Product}  \\ \hline
      Qubits & \textsf{ZZPhase} Depth & H$1$ & Emulator & Noiseless  \\ \hline
      10.00$\pm$0.00 &  262.00 $\pm$ 19.36     &  0.92$\pm$ 0.04   &  0.92$\pm$ 0.06  & 0.96$\pm$ 0.05      \\  \hline
  \end{tabular}
  \caption{Summary of the results of the execution of the reduced HHL circuits followed by the controlled-\textsf{SWAP} to obtain the inner product between the output state and the analytical solution loaded as a quantum state for five portfolio-optimization problem instances. It is displayed the number of qubits, the \textsf{ZZPhase} depth and the inner product obtained from the execution of the circuits on the trapped-ion device, its emulator with approximated noise and noiseless simulation. It is reported the median and the standard deviation across the circuits corresponding to the five problem instances considered.}
  \label{table:inner_product_harware}
  \end{table}

\section{Scaling up for Future Hardware Demonstrations}
\label{section:scale}
%\textcolor{red}{DH: maybe a more optimistic name?Like scaling up for future hardware. I think this section is about using the emulator with smaller error rates?}

In principle, the proposed Hybrid HHL$^{++}$ can be applied to any QLSP, including portfolio-optimization problems, for any given size. However, the noise present in the quantum near-term hardware may be a significant limitation on how big the linear system can be. Note that some conditions of the linear system can also prevent from obtaining speedups with HHL, for example the condition number. Refer to \cite{aaronson2015read}. We quantify how the noise in the device impacts the output of the semiclassical QPE and therefore the estimation of the eigenvalues to control on and the quality of the solution obtained with HHL, respectively. For this, we considered four portfolios with six assets, being the linear system $8 \times 8$. The characteristics of the circuits for the semiclassical QPE and HHL can been seen in Table \ref{table:big_portfolios}. In terms of the noise, we focus on the two-quit fault probability (also known as ``p2'') as the source of noise to study because the faulty two-qubit gates are currently the dominant source of error \cite{Pino_2021}. This is the probability of a fault occurring during a two-qubit gate. It corresponds to the asymmetric depolarizing probability of the device's fully entangling two-qubit gate. 

\begin{figure}[h!]%
    \centering
    \subfloat[\centering]{{\includegraphics[width= 7.8 cm]{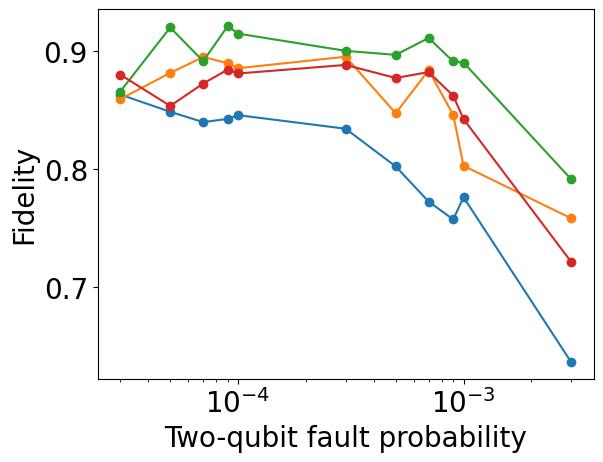} }}%
    \qquad
    \subfloat[\centering]{{\includegraphics[width= 8 cm]{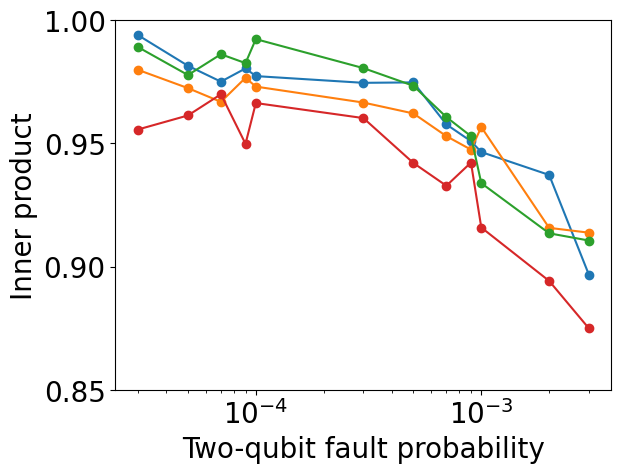} }}%
    \caption{Benchmark metrics as a function of the value of the two-qubit fault probability (``p2'') utilized to simulate the noise in the  emulator. Each color plots a different portfolio-optimization problem instance consisting of six assets ($8 \times 8$ linear systems of equations). The metric in (a) is the fidelity between the output distribution obtained from executing the semiclassical QPE circuit on emulator with the given noise and the distribution obtained in noiseless simulation. The metric in (b) is the inner product between the output quantum state of the HHL circuit executed on the emulator with the given level of noise and the analytical solution loaded as a quantum state.} %
    \label{fig:fidelity_big}
\end{figure}

We utilize the emulator and modify the value corresponding to this fault probability. Note that the default value is $1.38 \times 10^{-3}$, see Ref.~\cite{h1edatasheet}. We present results with both increased and decreased levels of noise For each value of the fault probability, we executed the semiclassical QPE circuits with $400$ shots on the emulator tuned with that value, and we calculated the fidelity with the output of the noiseless simulation. The results for the four problem instances considered can be seen in Figure \ref{fig:fidelity_big} (a) where we plot the fidelity as a function of the value of the two-qubit fault probability used to model the noise in the emulator. As expected, we observe that as the fault probability is reduced, the fidelity increases for the four problem instances considered. We note that this increase is not continuous but there are some slight ups and downs. However, the trend is clear. Additionally we observe that the difference in fidelity among the problem instances is mainly due to the eigenvalue distribution of each problem. 

We did the same experiment but for the HHL circuit controlling on the bitstrings of the eigenvalues calculated analytically up to four bits and with the angles calculated using these values. We executed these HHL circuits on the emulator with $2000$ shots using the different values of two-qubit fault probability, and we computed the inner product between the output quantum state of the HHL circuit and the analytical solution loaded as a quantum state. In practice one would do the steps in the proposed Hybrid HHL$^{++}$ algorithm and actually utilize the output distribution from the semiclassical QPE executed on hardware to get the estimates. However, as discussed before, for this problem size ($8 \times 8$), we have seen the the fidelity obtained on hardware is not good enough to allow us to use this output distribution. For this reason, for these experiments we used the real eigenvalues to estimate the four-bit bitstrings and to calculate the angles to rotate. We found the optimal scaling parameter following the proposed algorithm by performing the semiclassical QPE in noiseless simulator. In Fig.\ref{fig:fidelity_big} (b) we plot the inner product as a function of the level of noise for the problem instances considered. This plot gives us a sense of how the noise impacts the quality of the solution obtained and it allows us to get an idea of what future experiments on hardware would look like. We observe that as we reduce the two-qubit fault probability, the inner product improves for all the instances and for low levels of noise, it actually obtains very high values, demonstrating that the solution obtained with HHL is very close to the analytical solution. Note that when utilizing Hybrid HHL$^{++}$ we will not estimate the eigenvalues perfectly as we did for these experiments, and so we expect to see a decrease in the inner product. However, we believe that the rate at which the quality of the HHL solution improves as we decrease the noise will still hold and this analysis allows us to gain an understanding of how future demonstrations on hardware for larger problem sizes would look like.

\begin{table}[h!]
\normalsize
 \centering
  \begin{tabular}{l|*{4}{c}}
               &  Qubits & \textsf{ZZPhase} Depth & \\ \hline
      Semiclassical QPE &  4    &  356   &          \\  \hline
               %& Qubits & 5       & 6       & 7  \\ \hline
      HHL  &    12   &  964   &           \\
  \end{tabular}
  \caption{Circuit characteristics of the semiclassical QPE and HHL corresponding to portfolio-optimization instances with six assets ($8 \times 8$ systems). It is reported both the number of qubits and the \textsf{ZZPhase} depth of the circuits. Note that the values are the same for all the problem instances because the rotations are controlled on the set of the distinct analytical eigenvalues whose size is the same for all the instances.}
  \label{table:big_portfolios}
  \end{table}

\section{Discussion}
\label{sec:Conclusions}
In this work, we leverage newly hardware features with the development of algorithmic tools to make the HHL algorithm more feasible on NISQ  hardware. This results in a new application-oriented benchmark for small quantum hardware on quantum linear algebra tasks. This work represents a milestone towards the evaluation of quantum algorithms to tackle relevant use cases in quantum hardware available today. We improved on the Hybrid HHL introduced by Lee \etal in two ways. 

We developed an algorithm that optimizes the scaling parameter $\gamma$ for the Hamiltonian simulation required by phase estimation. We showed that the scaling of $\mathbf{A}$ by the $\gamma$, obtained with the algorithm, enabled resolving the relevant eigenvalues in the output distribution of phase estimation more accurately. Moreover, we introduce a heuristic method that reduces the complexity of the HHL circuit.

By levering newly-available hardware features we implemented a variant of the quantum phase estimation, named semiclassical QPE. This version of QPE is particularly suitable for NISQ devices since it addresses the scarcity of qubits by using just one ancilla for an arbitrary bit precision. It also tackles the fact that faulty two-qubit gates are currently the dominant source of error by replacing two-qubit gates with one-qubit gates controlled by classical bits. We have calculated with experimental results obtained on the Quantinuum H$1$ machine the fidelity between experimental and noiselessly simulated measurement distributions. First, for one problem instance we compared the results obtained with different number of bit precision %and for this particular problem size, which is the main parameter that imposes to circuit depth, we have determined that the best results are obtained with four bits precision. 
Then, we experimentally showed that the semiclassical variant achieves high fidelity for four-bit precision, over five different problem instances. This is achieved while reducing the number of qubits and the two-qubit gate count required. The main aim of this work is to allow for hardware demonstrations of quantum algorithms for tackling applications on near-term hardware. We do not claim any speedup with our method, it is a step forward towards evaluations on real hardware.

\section*{Acknowledgments}We thank Tony Uttley, Brian Neyenhuis and the rest of the Quantinuum Quantum Solutions team for assisting us on the execution of the experiments on the Quantinuum H-series trapped-ion devices.
We also thank Michele Mosca and his team at University of Waterloo for their insights, and Aram Harrow from Massachusetts Institute of Technology for his precious feedback.

\section*{Disclaimer}
This paper was prepared for information purposes by the Global Technology Applied Research group of JPMorgan Chase Bank, N.A.. This paper is not a product of the Research Department of JPMorgan Chase Bank, N.A. or its affiliates. Neither JPMorgan Chase Bank, N.A. nor any of its affiliates make any explicit or implied representation or warranty and none of them accept any liability in connection with this paper, including, but limited to, the completeness, accuracy, reliability of information contained herein and the potential legal, compliance, tax or accounting effects thereof. This document is not intended as investment research or investment advice, or a recommendation, offer or solicitation for the purchase or sale of any security, financial instrument, financial product or service, or to be used in any way for evaluating the merits of participating in any transaction.

\section*{Author contributions statement}

%P.N., R.Y. devised and implemented the optimization formulation. R.S., R.Y. implemented the quantum algorithms.  P.N., R.S., R.Y., D.H. performed the experiments. P.N., R.S., R.Y. analyzed the data.
R.Y. devised the algorithm idea. P.M. developed the algorithmic tools. P.M., R.Y., D.H. implemented the quantum algorithms and performed the experiments. P.M., R.Y., D.H. analyzed the data. M.P. supervised the project.  All authors contributed to the technical discussions, evaluation of the results, and the writing of the manuscript.

\section*{Additional Information}
The authors declare no competing interests. 

\section*{Data Availability Statement}
The data will be shared upon reasonable request. Please contact the corresponding author, Romina Yalovetzky.

\bibliography{sample}

\end{document}